\documentclass[a4paper,11pt]{article}
\usepackage{jcappub}
\usepackage{amsmath}
\usepackage{enumerate}
\usepackage{natbib}
\usepackage{url}
\usepackage{xcolor}
\usepackage{comment}
\usepackage{orcidlink}

\defcitealias{Umeda25}{U+25}
\defcitealias{Gagnon-Hartman26}{G-H+26}
\newcommand{\lya}{\mathrm{Ly}\alpha}

\title{Constraining reionization with persistent homology: a novel summary statistic for wide-field Lyman-$\alpha$ emitter observations}

\author[a]{Michael M. Wyatt\orcidlink{0009-0008-7752-7800}}
\affiliation[a]{Department of Physics and Astronomy, University of California, Los Angeles, CA, USA}

\author[b]{Nikolaos Triantafyllou\orcidlink{0009-0003-8609-4529}}
\affiliation[b]{Scuola Normale Superiore, 56126, Pisa, PI, Italy}

\author[a]{Steven R. Furlanetto\orcidlink{0000-0002-0658-1243}}

\author[c,d,e]{Hiroya Umeda\orcidlink{0009-0008-0167-5129}}
\affiliation[c]{Center for Computational Sciences, University of Tsukuba, Ten-nodai, 1-1-1 Tsukuba, Ibaraki 305-8577, Japan}
\affiliation[d]{Institute for Cosmic Ray Research,
The University of Tokyo,
5-1-5 Kashiwanoha, Kashiwa,
Chiba 277-8582, Japan}
\affiliation[e]{Department of Physics, Graduate School of Science, The University of Tokyo, 7-3-1 Hongo, Bunkyo, Tokyo 113-0033, Japan}

\author[b]{Samuel Gagnon-Hartman\orcidlink{0009-0006-7603-644X}}

\author[f,b]{Andrei Mesinger\orcidlink{0000-0003-3374-1772}}
\affiliation[f]{Department of Physics and Astronomy ``Ettore Majorana”, University of Catania, Catania, Italy}

\author[d,g,h,i]{Masami Ouchi\orcidlink{0000-0002-1049-6658}}
\affiliation[g]{National Astronomical Observatory of Japan, National Institutes of Natural Sciences, 2-21-1 Osawa, Mitaka, Tokyo 181-8588, Japan}
\affiliation[h]{Department of Astronomical Science, SOKENDAI (The Graduate University for Advanced Studies), 2-21-1 Osawa, Mitaka, Tokyo, 181-8588, Japan}
\affiliation[i]{Kavli Institute for the Physics and Mathematics of the Universe (WPI), University of Tokyo, Kashiwa, Chiba 277-8583, Japan}

\emailAdd{michael.wyatt@physics.ucla.edu}

\abstract{

The spatial clustering of Lyman-$\alpha$ emitters (LAEs) is one of the most promising observables for constraining reionization. In this work, we introduce persistent homology, a popular tool in topological data analysis, and specifically the Betti-0 curve as a summary statistic to characterize the clustering of LAEs. We compare the performance of the Betti-0 curve against the angular correlation function (ACF) and void probability function (VPF) by quantifying their ability to constrain the global neutral fraction $x_{\rm HI}$ using the Fisher information. We forward model observations of LAEs using \texttt{21cmFAST} by including $\lya$ emission as a post-processing step and account for cosmic variance uncertainties. We focus our analysis within the framework of the SILVERRUSH. catalog at $z=5.7$ and 6.6, and find that the Betti-0 curve performs as well as or better than the ACF and VPF. We also find that the statistics are most sensitive to different physical scales and thus probe complementary information, and as a consequence that simultaneously constraining all three statistics improves constraints by a factor of $\sim 2$ compared with the ACF alone, with predicted minimum uncertainties of $\sigma_{\rm min} \sim$ 0.03 and 0.07 at $x_{\rm HI} = 0.05$ and $z=5.7$ and $x_{\rm HI} = 0.20$ and $z=6.6$, respectively.
Our results show that the Betti-0 curve is a promising summary statistic to constrain reionization using LAE clustering and motivate further exploration of persistent homology for the same purpose in other wide-field surveys such as those made possible by the Nancy Grace Roman Telescope.
}

\begin{document}
\maketitle
\flushbottom

\section{Introduction}
Understanding and constraining the epoch of reionization (EoR) remains a major goal of modern cosmology and extragalactic astrophysics. Lyman-$\alpha$ emitters (LAEs) are one of the most promising probes of the epoch, as their visibility is modulated by the size of the ionized bubbles in which they are located due to the large optical depth of the neutral IGM. As a consequence, the severity of their spatial clustering is dependent on the morphology and timing of reionization, with LAEs appearing more clustered earlier in reionization, making summaries which are sensitive to this characteristic effective constraints of the EoR \citep{Furlanetto06, McQuinn07}.

There exists a variety of possible summary statistics to characterize the clustering of LAEs. Likely the most popular are two-point statistics such as the two-point correlation function, or its 2-dimensional equivalent, the angular correlation function (ACF), which has been measured using data from the Subaru telescope and used to constrain $x_{\rm HI}$ \citep{McQuinn07, Ouchi10, Sobacchi15, Ouchi18}. Most recently, it has been measured using the SILVERRUSH. catalog, which consists in part of 6124 and 2058 LAEs at $z = 5.7$ and 6.6 \citep{Kikuta23}, covering areas of 24.03 and 24.38 deg$^2$, respectively \citep{Umeda25}, which are selected using narrowband images from the Hyper Suprime-Cam Subaru Strategic Program (HSC-SSP; \cite{HSCSSP_design, HSCSSP_I, HSCSSP_II, HSCSSP_III}). Using the ACF, \cite{Umeda25} estimated the neutral hydrogen fraction as $x_{\rm HI}(z=5.7)=0.06^{+0.12}_{-0.03}$ and $x_{\rm HI}(z=6.6)=0.21^{+0.19}_{-0.14}$. 

However, the modulation of the visibility of LAEs is non-Gaussian. This is because the ionization field itself is approximately discrete as it is comprised of ionized bubbles with sharp boundaries. Because these ionized bubbles modulate the visibility and therefore the distribution of the LAEs, the resulting LAE distribution is also highly non-Gaussian. As such, the ACF is not sufficient to fully characterize their distribution, and so other statistics which capture at least some of this non-Gaussian information may allow for stronger constraints. Other statistics have been proposed and explored, such as counts-in-cells \citep{Mesinger08_early} or the related void probability function (VPF; \cite{Kashikawa06, McQuinn07, Perez21}). \cite{Gangolli21} find that the VPF can rule out a neutral Universe from a fully ionized one at a higher significance than the ACF in certain cases in a SILVERRUSH.-like survey. In particular, the authors find that the VPF outperforms the ACF in distinguishing $x_{\rm HI} = 0.3$ from $x_{\rm HI} = 0.0$ at $z=5.7$, but that the two statistics have similar constraining power at $x_{\rm HI} = 0.5$ and $z=6.6$.

In this work, we introduce a new category of summary statistics for inferences of reionization using LAEs: persistent homology. Persistent homology is a popular tool in topological data analysis which allows one to characterize topological features of a set of data points, opening up a range of statistical characterizations of such features, and as such is a natural fit for clustering-based analyses of discrete sources such as LAEs. While persistent homology has been applied to the topology of galaxy distributions for constraining cosmological parameters \citep{Heydenreich21, Yip24}, as well as the ionization field itself \citep{Elbers19,Giri21}, in this work we apply it to LAEs for the purpose of constraining reionization for the first time. In this work we explore one of the simplest statistics within the formalism of persistent homology, namely the ``Betti-0 curve,'' which tracks the number of groups of connected points --- known as ``components'' --- as a function of the grouping scale. Using forward models of LAEs based on $\texttt{21cmFASTv4}$ simulations \citep{Mesinger11, Murray20, Davies25}, we quantify the effectiveness of the Betti-0 curve against the ACF and VPF for constraining the timing of reionization through the global neutral fraction $x_{\rm HI}$ using Fisher information. 

This paper is organized as follows: in Section \ref{s:summary_stats} we provide an overview of the ACF and VPF before introducing persistent homology and the Betti-0 curve, and explore a qualitative comparison of the statistics. In Section \ref{s:forward_model} we describe our LAE forward model, including the empirical model for assigning $\lya$ emission, the reionization simulation, and LAE selection for reproducing the SILVERRUSH. catalog. In Section \ref{s:results} we present our results, and compare the constraining power of the Betti-0 curve against the ACF and VPF using Fisher information. We also show results including interlopers. In Section \ref{s:discussion} we explore why the three statistics are nondegenerate and probe complementary information, as well as show that the Betti-0 curve is as robust to astrophysical uncertainties as the other two statistics. In Section \ref{s:conclusion} we summarize our results and conclude.

Throughout this work we adopt the cosmology of \cite{planck_20} with $h = 0.6766$, $\Omega_{\rm m} = 0.30966$, and $\Omega_{\rm b} = 0.04897$.

\section{Summary statistics} \label{s:summary_stats}

In this section we review the ACF and VPF before providing a brief overview of persistent homology and describing the Betti-0 curve in Section \ref{ss:persistent_homology} and comparing the statistics qualitatively in Section \ref{ss:qual_compare}.

\subsection{ACF} 
To compute the angular correlation function,
we adopt the estimator of \cite{Landy93}:
\begin{equation}
    \omega(\theta) = \frac{DD(\theta)-2fDR(\theta)+f^2RR(\theta)}{f^2RR(\theta)},
\end{equation}
where $DD(\theta)$, $DR(\theta)$, and $RR(\theta)$ are counts of LAE-LAE, LAE-random, and random-random pairs at an angular separation of $\theta$, and $f=N_D/N_R$ is the ratio of the LAE to random number density.
We use \texttt{Corrfunc} to compute the pair counts \citep{Sinha20}. For each mock observation we count the number of observable LAEs and generate the same number of random points with a Poisson distribution over the same area at $z=5.7$, or 10 times as many at $z=6.6$. We ensure that the random points are constrained to the same pixel grid as the LAEs, so that they represent the same statistical distribution. 

\subsection{VPF}
The 2-dimensional void probability function is the fraction of randomly-placed circular regions which contain no LAEs as a function of the radius of the circles. Although the VPF can be computed at discrete radii, we choose to compute the VPF within bins, as is done with the ACF. For computational efficiency, we generate a k-d tree of the LAEs using \texttt{SciPy}'s \texttt{cKDTree} and then query the tree for the distance to the nearest LAE to each of our circle centers, $d_n$. For a radius bin between scales $r_{i}$ and $r_{i+1}$, the average value of the VPF for a random circle center $n$ is given by the fraction of radii within the bin which are smaller than the distance to the nearest neighbor:
\begin{equation}
(f_{0,i})_n=
    \begin{cases}
        0 & \text{if } d_n \leq r_i \\
        (d_n - r_i)/(r_{i+1} - r_i) & \text{if } r_i < d_n < r_{i+1} \\
        1 & \text{if } d_n \geq r_{i+1}.
    \end{cases}
\end{equation}
The value of the VPF in this bin is the mean across all $N_c$ randomly-placed circles: $f_{0,i} = N_c^{-1} \sum_{n=1}^{N_c} (f_{0,i})_n$. We consider 10,000 circles per slice of our simulation, or $\sim 2500$ deg$^{-2}$ (see Section \ref{ss:sim_obs} for details of our forward-models). The use of k-d trees for computing the VPF and other statistics is discussed in detail in \cite{Banerjee21}.

\begin{figure*}
    \centering
    \includegraphics[width=0.95\linewidth]{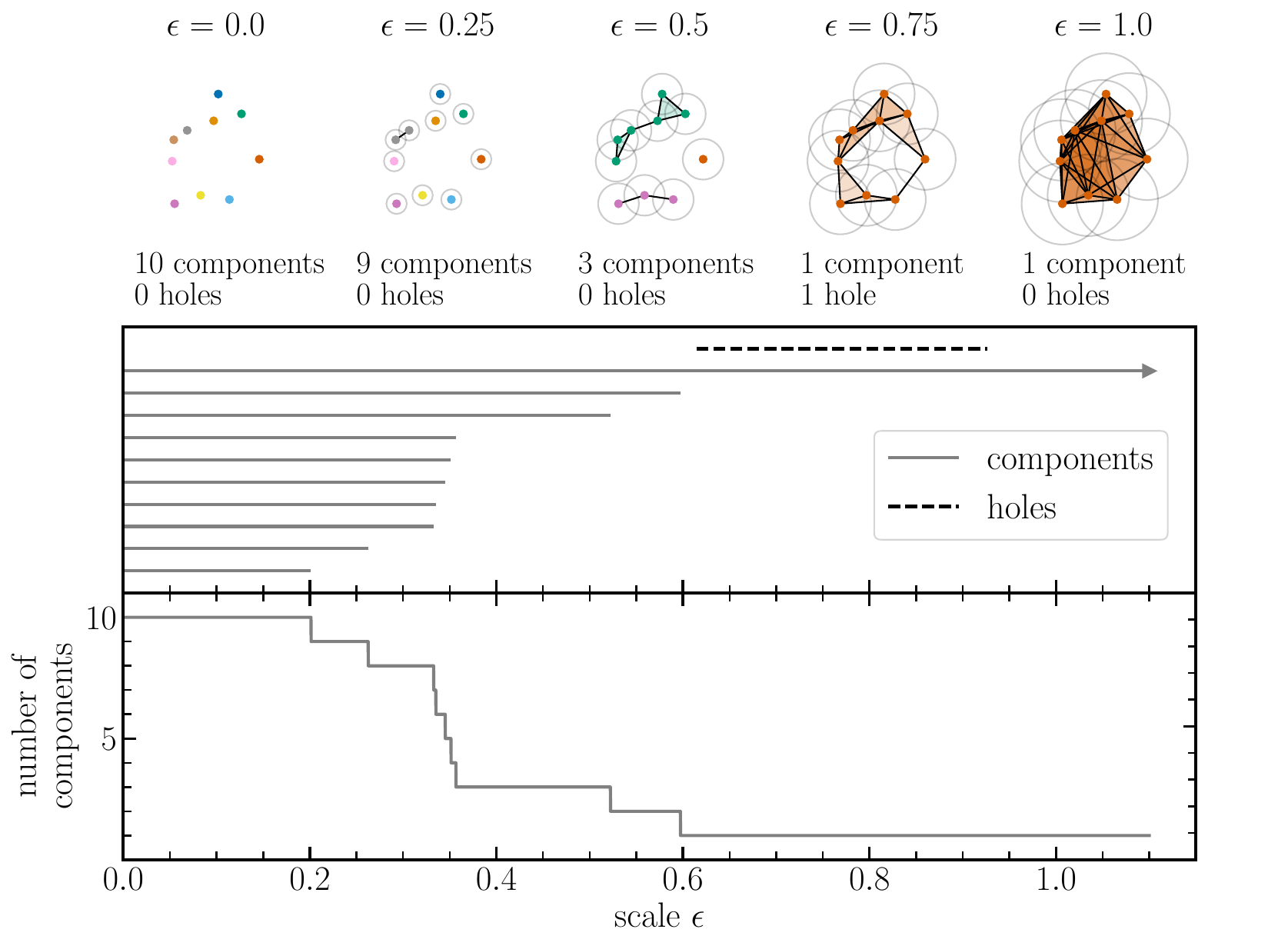}
    \caption{\textbf{Persistent homology for an example set of 10 points.} Top plots show the ``Vietoris--Rips filtration,'' which illustrates the features formed at several representative scales $\epsilon$. For this example, we consider only 0-simplices (points), 1-simplices (edges), and 2-simplices (triangles), and therefore also only $H_0$ features (connected components) and $H_1$ features (holes). Connected components are shown in common colors and circles are shown with radii equal to $\epsilon/2$ for the sake of visualization. The middle plot is the ``persistence barcode'' which shows the lifetimes of each component (solid lines) and hole (dashed line), and the bottom plot is the Betti-0 curve of the points, which shows the number of living $H_0$ features (components) as a function of scale. As there are 10 points, there are initially 10 components at $\epsilon=0$. As edges are formed, components ``die'' and the total number of components decreases monotonically until only a single component is left. This component has an infinite lifetime, as represented by a rightward facing arrow in the barcode. In this example, a single hole is formed and ``dies'' once it is filled in by triangles.}
    \label{fig:persistence_example}
\end{figure*}

\subsection{Persistent homology and Betti-$n$ curves}\label{ss:persistent_homology}
In this work we explore characterizing the clustering of LAEs using one of the simplest summary statistics within the persistent homology formalism: the ``Betti-0 curve.'' While a thorough review of persistent homology is not necessary for the current work, we will describe some of the pertinent features. Broadly, topological invariants allow a robust description of complex structure, and persistent homology is a method which allows one to characterize such features for a set of data points (such as a distribution of LAEs). Persistent homology is a popular tool in topological data analysis for a few reasons: (\textit{i}) its connection with algebraic topology opens up a rich set of tools for analysis, (\textit{ii}) it is deterministic for a given data set with no reliance on free parameters, (\textit{iii}) it is robust to small perturbations or noise in the data, and, (\textit{iv}) although backed by a deep mathematical formalism, the computation of persistent features in a 2-dimensional data set is conceptually and computationally straightforward. Though there exist many pedagogical resources on the subject, we refer the interested reader to the review of \cite{Otter17}, which we used as a reference for the current work.

We will describe persistent homology mostly through an example, where we consider only the 2-dimensional case in a Euclidean geometry (which is relevant to this work). Take Figure \ref{fig:persistence_example}, where we have several data points (0-simplices) in some 2-dimensional field. As a function of scale $\epsilon \geq 0$, we connect all pairs of points which are separated by a distance $\leq \epsilon$ with edges (1-simplices). This is represented in Figure \ref{fig:persistence_example} with circles of growing radius $\epsilon/2$, so that points are connected when their circles overlap. This method creates ``components'' of connected points, also known as $H_0$ features, or features of the zeroth homology group. A component is considered to be ``born'' on the scale at which it first comes into existence, and it ``dies'' on the scale at which it is combined with another existing component, where, by convention, the component that formed at the smaller scale is considered to persist. Given this definition, components are all ``born'' at a scale $\epsilon=0$, where there will be an equal number of components as there are data points, and the number of components decreases monotonically until it reaches 1 at a sufficiently large scale, at which point all points are connected. This procedure is analogous to using a ``friends-of-friends'' algorithm to group points together, but as a function of the linking length rather than at a fixed linking length (for examples of friends-of-friends analyses see e.g., \cite{Huchra82, Einasto84, Eke04, Berlind06}). 

One can also trace higher-dimensional simplices in a similar manner.\footnote{We note here that higher-dimensional simplices do \textit{not} require higher-dimensional data. Persistent homology is based only on the pairwise distance between points rather than on the dimension of the space in which they are embedded.} All sets of three points which are mutually separated by a distance $\leq \epsilon$ will form triangles (2-simplices; shown in Figure \ref{fig:persistence_example}), all sets of four points will form tetrahedra (3-simplices; not shown in the example), and so on for higher numbers of points. These higher-order simplices allow the formation of higher-order homology features, namely ``holes'' ($H_1$ features) which correspond to loops of edges that are not filled in by triangles, and ``voids'' ($H_2$ features) which correspond to closed shells of triangles that are not filled by tetrahedra. These higher order features are defined similarly to $H_0$ features but are ``born'' at nonzero scales and ``die'' on the scale at which they are completely filled in. 

One can track the \textit{persistence} of such features across scales; features which persist over a long range of scales are considered real features of the data, while short-lived ones can be attributed to noise (though we note that what constitutes long- or short-lived is not strictly defined). In the middle panel of Figure \ref{fig:persistence_example} we represent this with a so-called ``persistence barcode,'' which shows the birth and death scales for each individual feature. While useful for visualization, such a plot is difficult to apply directly to statistical inference. Instead, one often represents the persistence of data using ``Betti-$n$'' curves, which simply track the number of existing simplices of order $n$ as a function of $\epsilon$. Thus, the Betti-0 curve tracks the number of components. With the existence of several packages dedicated to computing persistent homology in \texttt{Python}, computing the Betti-0 (or higher order) curves is straightforward. 

Throughout this work, we normalize the Betti-0 curve by the total number of points and denote this quantity by $\beta_0$. With this normalization, the Betti-0 curve can be thought of as the \textit{fractional number of independent components}, or the \textit{inverse of the mean number of points per component}. Because we normalize each mock observation to have the same number of LAEs (see section \ref{ss:sim_obs}), this choice of normalization has no affect on our results, but facilitates comparison to the VPF and ACF. In an inference in which the number of LAEs is not fixed, there may be more optimal ways to normalize the Betti-0 curve, or indeed no normalization at all may be optimal.

Like the VPF and ACF, we choose to compute the Betti-0 curve as the average within bins, i.e.,
\begin{equation}
    \beta_{0,i} = (r_{i+1} - r_i)^{-1}\int_{r_i}^{r_{i+1}} \beta_0(\epsilon) d\epsilon.
\end{equation}
For convenience, we provide a short, publicly available \texttt{Python} function for computing the bin-averaged Betti-0 curve, which uses the \texttt{Ripser.py} package to compute the Vietoris-Rips filtration \citep{ripser_py, Ripser}.\footnote{\url{https://github.com/michaelmwyatt/betti0}}

Due to the relatively limited nature of the data currently available to LAE analyses, in this work we focus only on the Betti-0 curve as it is less noisy given the limited sample size, but we advocate for the continued exploration of persistent homology (and perhaps higher order features) for use in LAE statistics as observations continue to improve. We also note that, although we restrict ourselves to the case of narrowband observations, and thus a 2-dimensional application, 
persistent homology and the Betti-0 curve are just as applicable in 3-dimensions, if spectroscopic redshifts are available.

\begin{figure*}
    \centering
    \includegraphics[width=1.\linewidth]{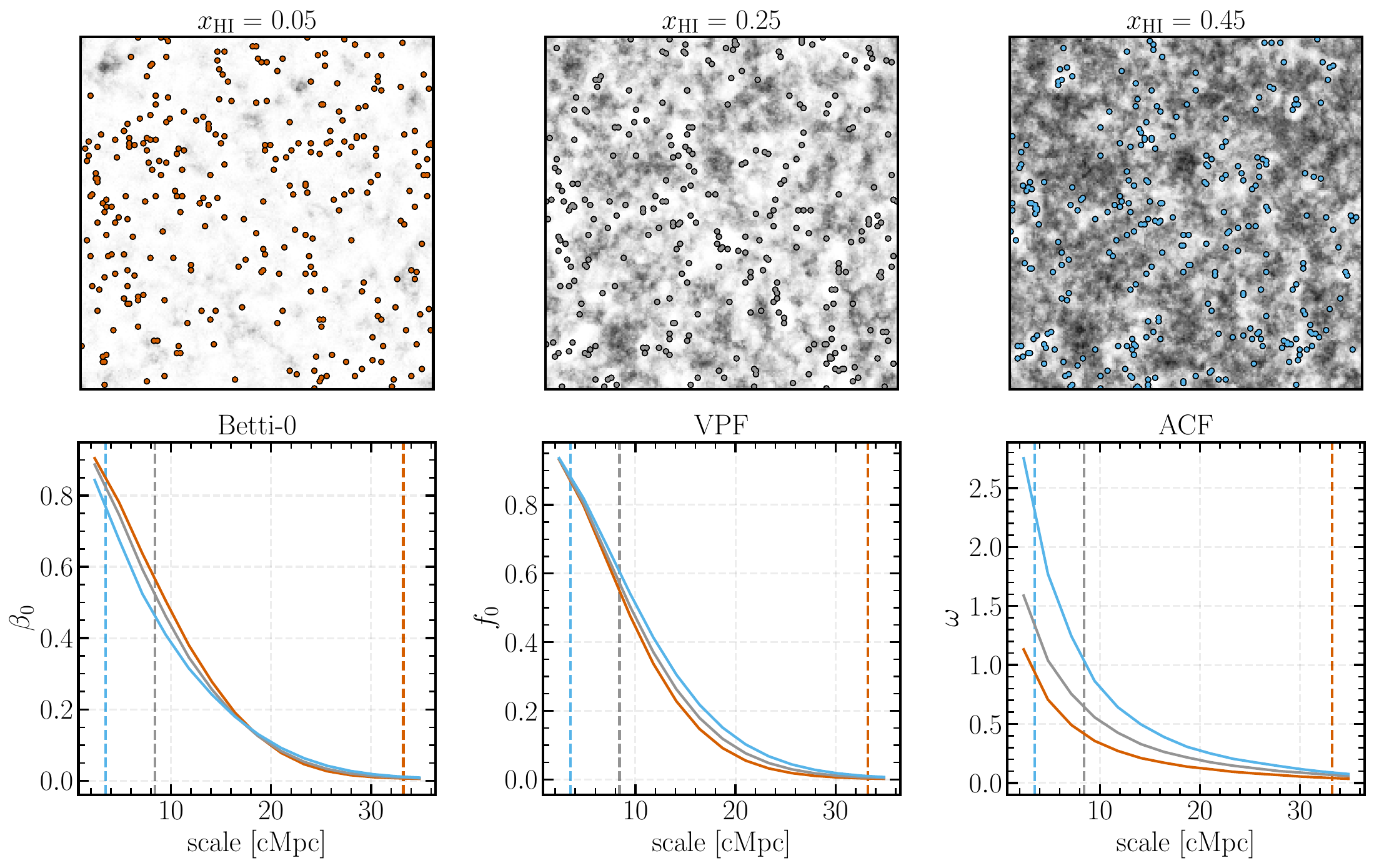}
    \caption{\textbf{The Betti-0 curve is sensitive to the change in the clustering of LAEs caused by a changing neutral fraction.} Top plots show representative slices of LAEs at $z=6.6$, with LAEs taken from the same underlying population, selected uniformly above an observed luminosity limit of $\log L_{\alpha, \rm att} > 42.6\,\text{erg}\,\text{s}^{-1}$ from a \texttt{21cmFAST} simulation and then randomly sampled to have a surface number density of $75 \, \text{deg}^{-2}$ (see Section \ref{s:forward_model} for details regarding the forward model). Grayscale shows the ionization field within each slice, with ionized regions shown in white. Slices are 300 $\times$ 300 $\times$ 42 cMpc ($\sim 4 \, \text{deg}^{2}$ in area). LAEs become visibly more clustered as $x_{\rm HI}$ increases and the ionized bubble size decreases, and all three statistics, including the Betti-0 curve, reflect the change in this clustering, shown in the bottom panels. The colors of each curve correspond to the same values of $x_{\rm HI}$ as the top panels, and each curve represents the average across 2000 simulations with the same astrophysical parameters but different initial conditions. Note that, although the ACF shows larger absolute changes with respect to the changing $x_{\rm HI}$, these changes combined with the covariance of the statistic ultimately determine the sensitivity --- we compare each statistic quantitatively in Section \ref{s:results}. In all of the bottom panels we illustrate the the characteristic ionized bubble size at each neutral fraction ($x_{\rm HI} = 0.45,0.25,$ and 0.05 from left to right) using vertical dashed lines, which we define using the `MFP' method. For the sake of visualization, in this example we have used a smaller luminosity threshold and chosen a larger number of LAEs than throughout the rest of the work.}
    \label{fig:stats_example}
\end{figure*}

\subsection{Qualitative comparison of statistics}\label{ss:qual_compare}

In order to provide some intuition for the Betti-0 curve compared with the other statistics, we explore an example computed on simulated data; we discuss our forward model in detail in Section \ref{s:forward_model}, but briefly describe some pertinent details here. 

In Figure \ref{fig:stats_example}, we show simulated LAEs at three different neutral fractions at $z=6.6$, along with the associated Betti-0 curve, VPF, and ACF. Each top panel represents a 300 $\times$ 300 $\times$ 42 cMpc slice from a single lightcone at various values for the global neutral fraction. LAEs are uniformly selected from the same population at $z=6.5$ with IGM-attenuated (observed) $\lya$ luminosities $\log L_{\alpha,\rm att} > 42.6 \, \text{erg} \,\text{s}^{-1}$. LAEs are then randomly sampled from this larger population so that each panel has a number density of 75 deg$^{-2}$.\footnote{Note that this number density is higher than the SILVERRUSH. sample for the sake of visualization. We present distributions more representative of the data in section \ref{s:results}.} We choose to compare statistics at a fixed number density so that any change in the clustering is due to the change in the ionization morphology, rather than the change in the number density itself. This normalization scheme is analogous to varying the duty cycle with $x_{\rm HI}$.

In the top panels we also show the $x_{\rm HI}$ field within each slice, where the shade of grey corresponds to the neutral fraction averaged along the thickness of the entire slice, with fully ionized regions shown in white. For reference, we indicate the characteristic ionized bubble scale for each of the three slices (i.e., each of the $x_{\rm HI}$ values) in all of the bottom panels with dashed lines. We measure ionized bubble sizes using the `mean free path' (MFP) method \citep{Mesinger07}. Rays are drawn with random directions uniformly distributed over the unit sphere, originating from random points within ionized regions. Once a voxel is reached with a neutral fraction greater than a threshold value (in this case, $x_{\rm HI, th} = 0.5$), the ray is terminated and its length $\ell$ is recorded. We account for periodic boundary conditions in the appropriate directions. As bubble sizes tend to be lognormally distributed, we define the characteristic bubble scale as the geometric mean of these lengths, $\log_{10} R_b \sim \langle \log_{10} \ell \rangle$.

The three statistics are shown in the bottom panels. For these examples, we have averaged each curve within 16 linearly-spaced bin edges between 1.5 and 36 cMpc (1 and 24 pixels in our simulations), and each curve represents the average over 2000 simulations with the same astrophysical parameters but different initial conditions (see Section \ref{s:forward_model} for more detail). To facilitate comparison, we also show the ACF as a function of separation in cMpc rather than angular separation and with a linearly-scaled (rather than log-scaled) ordinate.

The increased clustering is visually apparent as the neutral fraction increases, and all three statistics are distinct at different neutral fractions. The ACF behaves as expected: at sufficiently large scales, the clustering is unaffected by neutral fraction and the data number counts tend towards those of the randoms, and the ACF goes to $\sim 0$. At smaller scales, increased clustering causes an increase in the data counts compared with the randoms, and the curve increases with neutral fraction (e.g., \cite{McQuinn07}). Similarly, at sufficiently large scales the VPF tends to 0 in all cases (as infinitely large regions have no chance of being void), and increases with neutral fraction at smaller scales (e.g., \cite{Gangolli21}).

The Betti-0 curve has similar asymptotic behavior to the VPF, but at smaller scales it also exhibits a turnover, distinct from both the VPF and ACF.
Because the Betti-0 curve is monotonic and (in this case) normalized to have the same value at a scale of 0 regardless of clustering, it can be helpful to think about its behavior in terms of its slope. When ionized bubbles are smaller and thus the clustering is enhanced and the spacings between points shift to smaller values, the Betti-0 curve has a steeper slope at smaller values, where components are more likely to die as other points are encountered. Conversely, at larger scales, this trend reverses; since the points that would have been encountered at these scales were shifted to smaller ones, components are longer-lived, and the slope becomes flatter. We see this behavior reflected in Figure \ref{fig:stats_example}; compared with lower values of $x_{\rm HI}$, the Betti-0 curves corresponding to higher neutral fractions have initially steeper slopes which flatten at smaller scales, leading to the turnover that can be seen around 18 cMpc in the figure. At sufficiently large scales all curves become 1 as all components have died and only a single one remains. 

Although the ACF shows larger absolute changes with a changing $x_{\rm HI}$, we remind the reader both that Figure \ref{fig:stats_example} does not include error bars, and that the sensitivity of a given statistic is ultimately determined by both he change in its mean and its covariance.
As such, these trends motivate a more quantified exploration of the sensitivity of the Betti-0 curve to $x_{\rm HI}$. Before doing so, we explain our process for forward-modeling $\lya$ emitters and creating mock observations for this purpose.

\section{Forward-modeling Lyman-$\alpha$ emitter observations}\label{s:forward_model}
The basis of our Ly$\alpha$ emitter forward modeling is a suite of simulations run using the state-of-the-art semi-numerical simulation \texttt{21cmFASTv4} \citep{Mesinger11, Murray20, Davies25}. These simulations provide both large-scale realistic ionization field distributions across a wide range of redshifts and also discrete halo locations along with associated properties relevant to computing the UV-magnitude of galaxies within these halos. As a post-processing step, we assign galaxies in the simulation an emergent Ly$\alpha$ luminosity $L_{\alpha}$ (i.e., the luminosity leaving the interstellar/circumgalactic medium and entering into the IGM, prior to any IGM attenuation) and velocity offset of Ly$\alpha$ compared with the rest frame of the galaxy $\Delta v$ using an empirical model which is a function 
of the UV-magnitude of the galaxy $M_{\rm UV}$.
We then account for the attenuation of Ly$\alpha$ due to intervening neutral gas in the IGM. To approximate the selection criteria of the SILVERRUSH. survey, we apply a uniform $\lya$ luminosity cut on these attenuated luminosities.  

In the following sections we first describe the model that we use for assigning emergent Ly$\alpha$ luminosities and velocity offsets (Section \ref{ss:LAE_models}), then the relevant details of the simulation suite (Section \ref{ss:database}), before finally describing the selection criteria and mock observations (Section \ref{ss:sim_obs}). 

\subsection{Emergent Ly$\alpha$ luminosities}\label{ss:LAE_models}

Assigning emergent Ly$\alpha$ emission to halos has been recognized as an important step in forward modeling LAE observations, and is often parameterized as an exponential, log-normal, or normal distribution plus a delta function conditional on the UV emission of the galaxy \citep{Dijkstra12, Treu12, Schenker14, Mason18}. In this work,
we adopt the method of \cite{Gagnon-Hartman26} (hereafter \citetalias{Gagnon-Hartman26}), where the authors model the conditional probability of LAE properties as a multivariate Gaussian which is a function of $M_{\rm UV}$: $P(\boldsymbol{x}_{\alpha} | M_{\rm UV})$, where $\boldsymbol{x}_{\alpha}$ is the vector of LAE properties:
\begin{equation}
    \boldsymbol{x}_{\alpha} = \{ \log_{10} L_{\alpha}, \; \Delta v, \; \log_{10} L_{\rm H\alpha} \},
\end{equation}
and $L_{\rm H\alpha}$ is the H$\alpha$ luminosity of the galaxy, which we do not need explicitly for the current work. The authors fit to $z \sim 5$ LAEs presented in \cite{Tang24} from the MUSE-Wide and MUSE-Deep fields with complementary observations from the JWST FRESCO program (MUSE-Wide \cite{Urrutia19}; MUSE-Deep \cite{Bacon17,Bacon23}; JWST FRESCO \cite{Oesch23}), for which the three properties above are reported. As is commonly done, we assume no redshift evolution in the conditional probability and so apply the fit as-is to our higher-redshift simulations. The authors carefully forward-model the selection effects of the sample when inferring the underlying probability distribution $P(\boldsymbol{x}_{\rm Ly\alpha} | M_{\rm UV})$ in order to remove the bias that is imposed by ignoring such a step, which should allow the LAE forward-modeling to be more accurate when reproducing surveys with different selection criteria than those used to fit the model originally.

\subsection{Simulations} \label{ss:database}

To produce realistic LAE distributions along with reionization morphology we utilize a suite of simulations created using \texttt{21cmFASTv4} (Triantafyllou et al. in prep.).\footnote{\url{https://github.com/21cmfast/21cmFAST/tree/2025-database-runs}}
We use a portion of the suite which is comprised of 2,000 runs that were
created using identical cosmological and astrophysical
parameter values but with randomized starting seeds. The parameter values are the fiducial values of \texttt{21cmFASTv4} (using the discrete halo sampler of \cite{Davies25} down to a mass of $M_h=10^9\,M_\odot$) while including both atomic cooling galaxies and molecular cooling galaxies when calculating inhomogeneous radiation fields (\cite{Qin20,Munoz22}; for more details of the simulation suite see \cite{Triantafyllou26}).

Each run produces a lightcone of cross sectional area $300 \times 300$ cMpc$^2$ comprised of cubical voxels with a side length of 1.5 cMpc across a redshift range of $z=35$ to $z=5$. For every voxel $i$, we have the value of its redshift $z_i$, overdensity $\delta_i$, and neutral fraction $x_{{\rm HI},i}$. Each run also produces six $300^3$ cMpc$^3$ coeval cubes at $z=9,8,7,6.5,6$, and $5.5$ with individual halo locations and masses for halos with $M_h > 10^{10} M_\odot$. Halos are assigned stellar masses through a log-normal stellar-to-halo mass relation, and subsequently star formation rates (SFRs) through the star forming main sequence, which is also assumed to be log-normal. These relationships are motivated by both hydrodynamical simulations and observations \citep{Nikolic24}. We assume that each halo contains a single galaxy. When forward-modeling an observation at a given redshift (in this work, $z=5.7$ and $z=6.6$) across a range of $x_{\rm HI}$ values, we populate slices of the lightcone using halos from the same coeval cube with the closest available redshift ($z=5.5$ and $z=6.5$) at various values for $x_{\rm HI}$, disregarding the associated redshift value of the lightcone itself. This allows us to isolate the effect of the changing ionization morphology from an evolving galaxy population. This is analogous to varying the ionizing efficiency to achieve the desired $x_{\rm HI}$ value (e.g., \cite{Mesinger08_early, Davies22, Wyatt26}), and is an appropriate approximation as reionization models tend to be more sensitive to $x_{\rm HI}$ than to redshift \citep{Furlanetto04, McQuinn07b}. We note that, due to the alignment of the coeval cubes with the lightcone, that this does not necessarily mean that we are considering the exact same halos in each slice of a given lightcone, but rather that halos are from the same statistical population for a given redshift.

To assign UV-magnitudes to each halo, we assume that the SFR of a galaxy is proportional to its rest-frame UV continuum luminosity density at 1500 {\AA} (e.g., \cite{Madau14}): 
\begin{equation}
    L_{\rm{1500}, \nu}(\dot{\rho}_\star) = \dot{\rho}_\star / \mathcal{K}_{\rm{UV}}, 
\end{equation}
where $\mathcal{K}_{\rm{UV}} = 1.15 \times 10^{-28} M_\odot$ yr$^{-1}$ / erg s$^{-1}$ Hz$^{-1}$ assumes a continuous mode star formation with a Salpeter IMF (e.g., \cite{Sun16}). 

To account for the attenuation of Ly$\alpha$ due to the IGM, we use the same approach as \cite{Mesinger08_damping} and compute the total line center Ly$\alpha$ optical depth for each galaxy towards lower redshifts by summing the contribution to the optical depth in each pixel along the line-of-sight to a distance of 75 cMpc using the approximation of \cite{Miralda-Escude98}:

\begin{equation}\label{eq:tau}
\tau = \frac{\tau_{\mathrm{GP}} R_\alpha}{\pi} \sum_i (1 + \delta_i) x_{\mathrm{HI},i}
\left( \frac{1 + z_{b,i}}{1 + z_{\rm eff}} \right)^{3/2}
\left[
I\left( \frac{1 + z_{b,i}}{1 + z_{\rm eff}} \right)
-
I\left( \frac{1 + z_{e,i}}{1 + z_{\rm eff}} \right)
\right]
\end{equation}

\begin{equation}
I(x) \equiv \frac{x^{9/2}}{1 - x}
+ \frac{9}{7} x^{7/2}
+ \frac{9}{5} x^{5/2}
+ 3 x^{3/2}
+ 9 x^{1/2}
- \ln \left| \frac{1 + x^{1/2}}{1 - x^{1/2}} \right|
\end{equation}

\begin{equation}
    \tau_{\mathrm{GP}}(z) \approx 4.9 \times 10^5 \left( \frac{\Omega_m h^2}{0.13} \right)^{-1/2} \left( \frac{\Omega_b h^2}{0.02} \right) \left( \frac{1+z_s}{7} \right)^{3/2},
\end{equation}

where $\tau_{\mathrm{GP}}$ is the Gunn-Peterson optical depth of a fully neutral IGM (\cite{Gunn65, Fan06}; note that the dependence on the neutral fraction has been moved to inside the sum), $\nu_\alpha~=~2.47 \times 10^{15}~\mathrm{Hz}$ is the rest frequency of the $\lya$ line, and $R_\alpha = \Lambda/4\pi \nu_\alpha$, $\Lambda = 6.25 \times 10^8~\mathrm{s}^{-1}$ is the decay constant for the $\lya$ resonance. $\delta_i$ and $x_{{\rm HI},i}$ are the overdensity and neutral fraction of each pixel, and $z_{b,i}$ and $z_{e,i}$ are the redshift at the \textit{beginning} and \textit{end} of each pixel (from the perspective of the line of sight). If the pixels along a line of sight have redshifts $z_i$, where $i$ runs along the length of the line-of-sight, then we assign $z_{b,i} = z_i$ and $z_{e,i} = z_{i+1}$. $z_{\rm eff}$ is the redshift of the Ly$\alpha$ line accounting for the velocity offset; in terms of the cosmological redshift of the galaxy, it is $z_{\rm eff} = z + (\Delta v/{\rm c})(1+z)$. We do not account for peculiar velocities when computing the optical depth. The value of the attenuated (i.e., observed) $\lya$ luminosity is given by

\begin{equation}
    L_{\alpha, \rm att} = L_{\alpha} \mathrm{e}^{-\tau}.
\end{equation}
In order to maintain consistency with the placement of halos within the lightcone, as mentioned above, we do not use the redshifts from the lightcone itself. Instead, we find the closest redshift in the lightcone to the desired redshift for the mock observation (either $z=5.7$ or 6.6), and assign this redshift to the center of our slice of width 80 voxels, and assign redshifts to other voxels relative to this value. We then choose a smaller sub-slice to match the FWHM of the narrowband filter of the HSC-SSP observations which is closest to the desired $x_{\rm HI}$ value (see the following section).

\subsection{Mock observations}\label{ss:sim_obs}
To provide a practical comparison of the ability of the Betti-0 curve to constrain $x_{\rm HI}$ against existing summary statistics, we wish to reproduce a realistic survey scenario. As such, we turn to the SILVERRUSH. sample as one of the most effective existing surveys for LAE constraints. While the exact performance of each statistic is dependent on the survey details, we expect the Betti-0 curve to remain a promising summary statistic for all similar wide-area LAE surveys. 

With the goal of reproducing the LAE selection from SILVERRUSH. XIII. \citep{Kikuta23}, we follow a similar strategy as SILVERRUSH. XIV. (\cite{Umeda25}; hereafter \citetalias{Umeda25}). We first describe the procedure in broad terms and then discuss each point in detail below. After populating the lightcones with the appropriate halo populations, assigning $L_\alpha$, accounting for the IGM attenuation (Section \ref{ss:database}), and selecting broad slices at fixed values for $x_{\rm HI}$ values, we select thin slices to match the full width at half-maximum (FWHM) of the narrowband filters, and combine enough slices to reproduce the full survey area. We then perform a uniform luminosity cut on the attenuated $\lya$ luminosity. Although the LAE selection criteria from SILVERRUSH. result in non-uniform coverage, sufficiently bright LAEs (with narrowband magnitudes brighter than 25 mag) can be selected homogeneously across the entire observation --- it is to this sub-population that we compare our results.

We approximate the narrowband selection of SILVERRUSH. as a uniform selection in redshift using the FWHM of the transmission curves, $z = 5.72 \pm 0.046$ and $z = 6.58 \pm 0.056$ \citep{Kikuta23}. To select slices of our lightcones, we convert these to comoving distance and find $\sim 42$ cMpc, or $\sim 28$ voxels in the simulations. As mentioned above, for the mock observations we select these slices from the lightcones at discrete values of $x_{\rm HI}$ relevant to each redshift. For the Fisher information analyses in section \ref{s:results} we consider slices spaced at intervals of $\Delta x_{\rm HI} = 0.05$ between $x_{\rm HI} = 0.05$ and 0.30 at $z=5.7$ and $x_{\rm HI} = 0.05$ and 0.50 at $z=6.6$.
Each lightcone has a cross-sectional area of 
around 4.3 and 3.9 deg$^2$ at $z = 5.7$ and 6.6, while the total survey areas of SILVERRUSH. are 24.03 and 24.38 deg$^2$ (\citetalias{Umeda25}). 
To reproduce the survey area, we cannot simply tile multiple slices as they would not be properly correlated. We also avoid tiling a single slice using periodic boundary conditions, as this would not increase the information contained within a single slice and would only serve to amplify any noise that exists within a given slice. In addition, the SILVERRUSH. data itself is not a single field, but instead comprised of 4 distinct fields. As such, to create a single mock observation we average the statistics across enough slices to approximately reproduce the desired total survey area, which in this case is 6, meaning that we have $2000/6 \approx 333$ independent mock observations at each redshift and $x_{\rm HI}$. Because each field is slightly larger than the area of our slices, we cannot access the largest available scales of the observations. However, as we show in Section \ref{s:results}, the summary statistics considered in this work are most sensitive at scales smaller than our lightcone width given the observed LAE number densities. 

\begin{figure}
    \centering
    \includegraphics[width=0.8\linewidth]{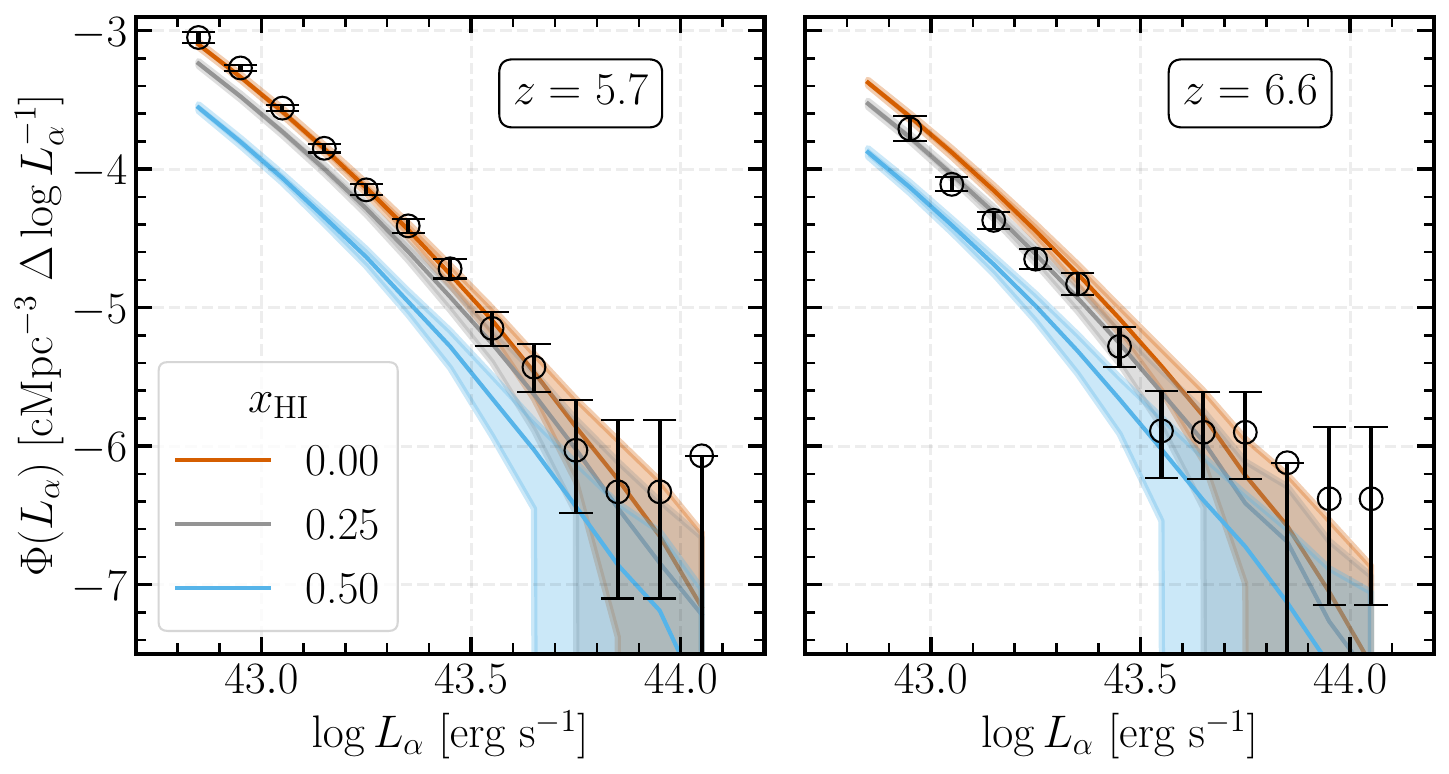}
    \caption{\textbf{Ly$\boldsymbol{\alpha}$ luminosity functions created with \texttt{21cmFAST} in combination with the \citetalias{Gagnon-Hartman26} model are consistent with those measured in SILVERRUSH. XIV. (\citetalias{Umeda25})}. Curves show the $\lya$ LFs computed using the method outlined in Section \ref{ss:sim_obs}, colors show results for global neutral fractions of $x_{\rm HI} = 0.00,$ 0.25, and 0.50 (from top to bottom), and error bars represent cosmic variance after accounting for the total survey area. Circles and error bars show the results from \citetalias{Umeda25}. Panels from top to bottom show observations at $z=5.7$ and 6.6. At $z=5.7$, where \citetalias{Umeda25} find a low neutral fraction of $x_{\rm HI}\sim0.05$ based on constraints using the LF and ACF, the unattenuated LF provides an approximate match to the observations. At $z=6.6$, a nonzero $x_{\rm HI}$ is required to match the data, implying that reionization is still ongoing on this redshift, also consistent with the results of \citetalias{Umeda25}.}
    \label{fig:LAE_LF}
\end{figure}

Galaxies with narrowband magnitudes brighter than 25 mag can be homogeneously sampled across the entire SILVERRUSH. survey at $z=5.7$ and 6.6, which corresponds to $\log L_{\alpha, \rm att}/$erg~s$^{-1}=42.7$ and 42.8, respectively, assuming an equivalent width of $EW \approx 20 \, \text{\AA}$ (\citetalias{Umeda25}). Within each slice, we select only LAEs with attenuated $\lya$ luminosities above these threshold values. 25 mag corresponds to approximately 50\% of the area-weighted detection completeness of the survey at these redshifts, and so uniformly selecting all galaxies above this limit within our simulation over-selects LAEs compared with the observations. For consistency with the observations, we then randomly sample these luminosity-selected LAEs so that each slice matches the average number density of the catalog. This normalization scheme is analogous to varying the duty cycle with $x_{\rm HI}$ and within each slice. The total number of LAEs in this uniformly-selected sample are 3170 and 321 at $z=5.7$ and $6.5$, respectively (\citetalias{Umeda25}),\footnote{The LAE sample sizes quoted here correct typographical errors in the text of \citetalias{Umeda25}; the analysis and results in \citetalias{Umeda25} are unaffected.} resulting in 567 and 53 LAEs in each of our simulation slices. Were the number of LAEs not normalized, the dominant source of sensitivity of each statistic would be the decrease in the number of observed LAEs as the neutral fraction rises. This change is highly sensitive to the choice of astrophysical parameters in the model \citep{Mesinger08_early}, and so we normalize the number density in order to isolate the sensitivity of each statistic to the changing morphology of the ionization field from the change in the underlying galaxy population itself. 

Both the stochastic grouping of slices and the stochastic random sampling of LAEs can affect our results slightly. As such, we repeat the random sampling of each of our 2,000 slices at each $x_{\rm HI}$ 50 times before making 100,000 groups of 6 randomly chosen slices, which we find is enough to ensure the convergence of our results. 

\subsection{$\lya$ luminosity function}
In Figure \ref{fig:LAE_LF} we show that the $\lya$ luminosity functions (LFs) resulting from the combination of \texttt{21cmFAST} and the \citetalias{Gagnon-Hartman26} $\lya$ model are consistent with those measured in SILVERRUSH. XIV. (\citetalias{Umeda25}). We show results at $z=5.7$ and 6.6 for a variety of $x_{\rm HI}$ values. Using the LF and ACF, \citetalias{Umeda25} find that reionization is nearly complete by $z=5.7$, and so it is reassuring that the data fall close to the LF at $x_{\rm HI} = 0$. On the other hand, \citetalias{Umeda25} find that reionization is still ongoing on $z=6.6$ --- consistent with this result, we find that the LF at $x_{\rm HI} = 0$ lies above the data at this redshift.

\begin{figure*}
    \centering
    \includegraphics[width=1.\linewidth]{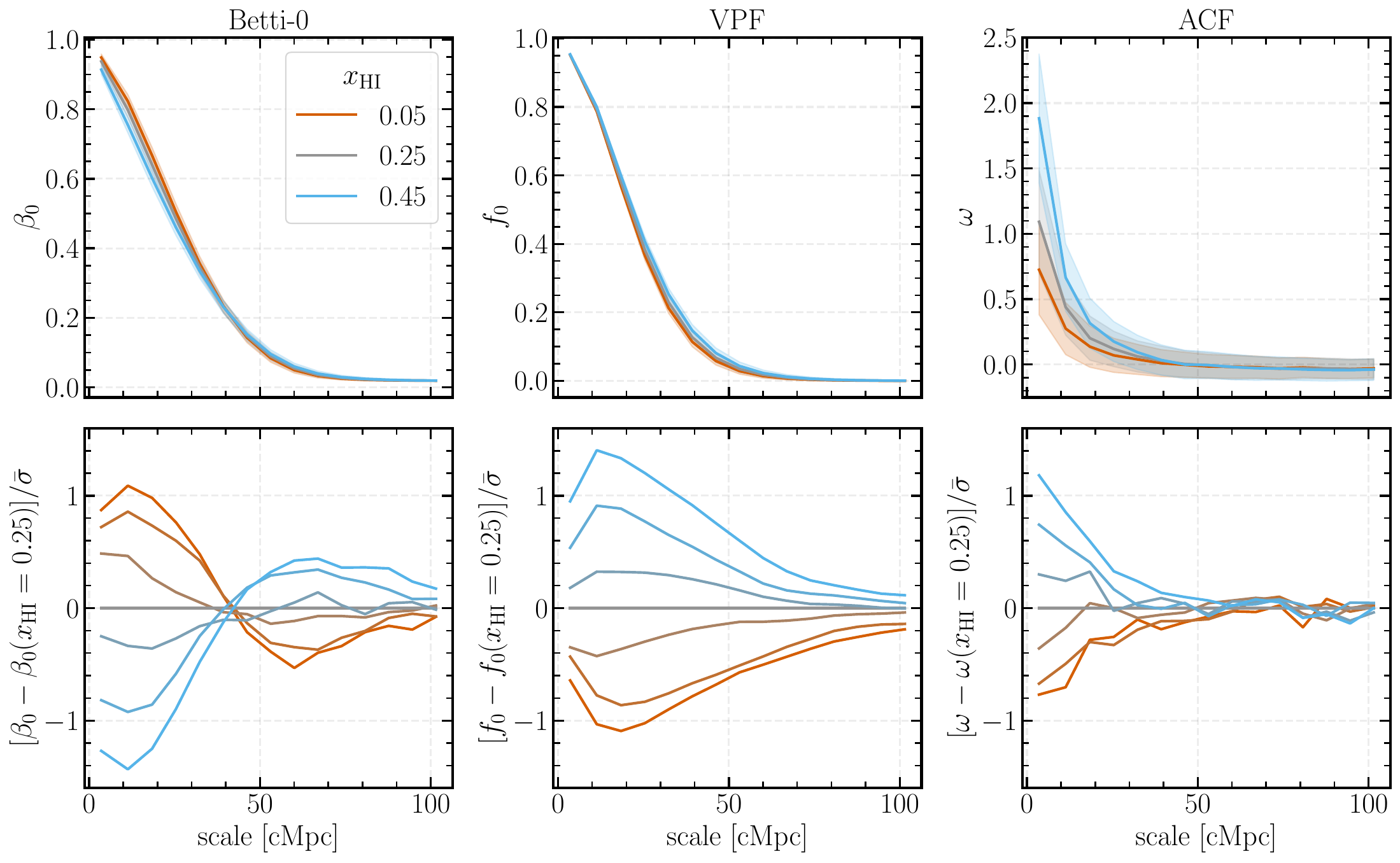}
    \caption{\textbf{The Betti-0 curve shows similar differences between $x_{\rm HI}$ values as the VPF and ACF, as a fraction of cosmic variance uncertainties.} Top panels show all three statistics computed at three global neutral fractions with uncertainties due to cosmic variance. The same qualitative trends discussed in Figure \ref{fig:stats_example} are present. Bottom panels show the difference of each curve at $x_{\rm HI} = \{0.0, 0.1, 0.2, 0.3, 0.4, 0.5\}$ and at $x_{\rm HI} = 0.25$ (colors from orange to blue; curves top to bottom at a scale of 10 cMpc for the Betti-0 curve, and bottom to top for the VPF and ACF), divided by the root mean square of the standard deviations at both neutral fractions, $\bar{\sigma} = \sqrt{[\sigma^2 + \sigma^{2}(x_{\rm HI} = 0.25)]/2}$. The magnitudes of the fractional differences of the Betti-0 curve and VPF are similar across the scales shown, while those of the ACF tend to be smaller. Differences at larger neutral fractions are slightly larger in magnitude, consistent with an increased sensitivity of the statistics at these values of $x_{\rm HI}$. }
    \label{fig:stats_comparison}
\end{figure*}

\section{Results}\label{s:results}

In the top panels of Figure \ref{fig:stats_comparison} we show results for all three summary statistics from our simulations at $z=6.6$. We plot the mean values for each statistic along with $1\sigma$ error bars, which represent cosmic variance uncertainties. The Betti-0 curve shows a clear sensitivity to $x_{\rm HI}$ and follows the same qualitative trends as described in Figure \ref{fig:stats_example}. As expected, the separation of curves for all statistics is slightly larger at higher values of $x_{\rm HI}$, reflecting the increased sensitivity of clustering at larger neutral fractions when the bubbles are smaller and the damping wing imprint is on average stronger. This effect is marginal here as we consider only $x_{\rm HI} \leq 0.5$, above which it becomes more pronounced. \citep{Mesinger08_early}. We note that we do not consider periodic boundary conditions when computing the statistics.

To make the sensitivity clearer, in the bottom panels of Figure \ref{fig:stats_comparison} we show the difference between each summary statistic at various neutral fractions and its value at $x_{\rm HI} = 0.25$, divided by the root mean square of the standard deviations at both neutral fractions.
From these panels we can see that the largest separation of the Betti-0 curve occurs at small scales, and that this diminishes towards larger scales. There is a sign flip at intermediate scales corresponding to the crossover of the curves visible in the top panels. Larger separations at larger values of $x_{\rm HI}$ again reflect the increased sensitivity of the statistic at larger neutral fractions. The separations are similar in magnitude for the Betti-0 and VPF curves across the scales shown, and both show larger separations than the ACF. 

\begin{figure*}
    \centering
    \includegraphics[width=1.\linewidth]{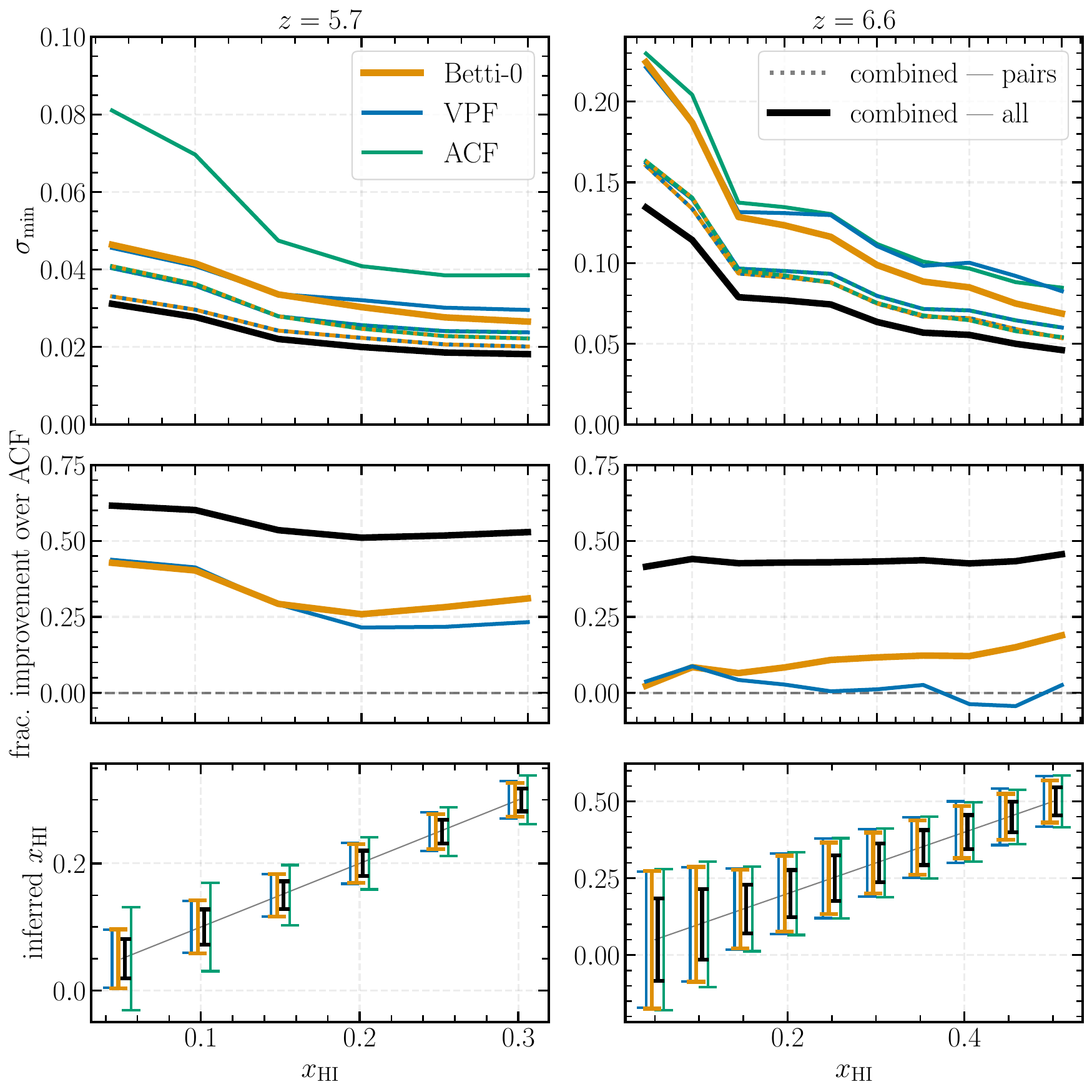}
    \caption{\textbf{The Betti-0 curve reduces uncertainties on constraints of $\boldsymbol{x_{\rm HI}}$ by up to 40\% compared with the ACF in a SILVERRUSH.-like survey, and performs at least as well as, and in many cases better than, the VPF for constraining $\boldsymbol{x_{\rm HI}}$. Simultaneous constraints of all three statistics  outperform  any individual or pair of statistics, and improve constraints by a factor of $\boldsymbol{\sim} \boldsymbol{2}$ compared with the ACF.} Top panels show the minimum uncertainty computed using the Fisher information $\sigma_{\rm min} = 1/\sqrt{F}$ for the Betti-0 curve (orange), the VPF (blue), and the ACF (green) at $z=5.7$ (left panel) and 6.6 (right). Solid lines indicate the constraints possible if each statistic is considered individually, while dotted lines show $\sigma_{\rm min}$ for simultaneous constraints of each pair of statistics (with matching colors), and black lines show $\sigma_{\rm min}$ for simultaneous constraints of all three statistics. Middle panels show the fractional improvement of the Betti-0 curve and VPF as well as the combined statistics compared with the ACF $(\sigma_{\rm min}^{\rm ACF}-\sigma_{\rm min})/\sigma_{\rm min}^{\rm ACF}$. Bottom panels show the inferred values for $x_{\rm HI}$ for each statistic with $1\sigma$ uncertainties assuming the mean is recovered exactly. In these panels unphysical values of $x_{\rm HI}$ are produced when the uncertainties extend below $x_{\rm HI} = 0.0. $ We compute each statistic within 10 linearly-spaced bin edges between 1.5 and 18.6 cMpc (1 and 12.4 pixels in our simulations) at $z=5.7$ and 1.5 and 63.6 cMpc (1 and 42.4 pixels) at $z=6.6$.}
    \label{fig:fisher_standard}
\end{figure*}

\subsection{Fisher information}\label{ss:fisher}

In order to quantitatively compare the summary statistics we use the Fisher information $F$, which provides a measure of the amount of information that some observable $\boldsymbol{s}$ carries about a parameter $\varphi$. Specifically, it provides the maximum possible constraining power of $\boldsymbol{s}$ on the parameter $\varphi$ through the Cram\'er-Rao bound in the form of the minimum possible standard deviation \citep{Rao1992, Cramer1999, Prelogovic24}: $\sigma(\varphi) \geq 1/\sqrt{F}$.\footnote{We note that the Fisher information here is a scalar rather than a matrix as we consider only the sensitivity to a scalar parameter $\varphi = x_{\rm HI}$.}

In this work, $\boldsymbol{s}$ is the vector of values in our summary statistic (for example, the value of the Betti-0 curve at various values of $\epsilon$), and $\varphi$ is the global neutral fraction $x_{\rm HI}$ --- for clarity we will use $\varphi = x_{\rm HI}$ hereafter. We assume that the data vector $\boldsymbol{s}$ is approximately multivariate normal (see Appendix \ref{a:gaussianity}), and that the covariance is not a strong function of $x_{\rm HI}$, which allows us to approximate the Fisher information as:

\begin{equation}\label{eq:fisher}
    F = \frac{\partial \boldsymbol{\mu}^T }{d x_{\rm HI}} \boldsymbol{\Sigma}^{-1} \frac{\partial \boldsymbol{\mu}}{d x_{\rm HI}}.
\end{equation}
As described in Section \ref{ss:sim_obs}, we compute the mean and covariance of a given statistic $\boldsymbol{s}$ at $x_{\rm HI}$ across $N~=~100,000$ groups of 6 simulation slices using the maximum likelihood estimator:
\begin{align}
    \boldsymbol{\mu}(x_{\rm HI}) &= \frac{1}{N}\sum_{i=1}^N \boldsymbol{s}(x_{\rm HI}) \\
    \boldsymbol{\Sigma}(x_{\rm HI}) &= \frac{1}{N}\sum_{i=1}^N (\boldsymbol{s}_i(x_{\rm HI}) - \boldsymbol{\mu}(x_{\rm HI}))(\boldsymbol{s}_i(x_{\rm HI}) - \boldsymbol{\mu}(x_{\rm HI}))^T.
\end{align}
There is a known inherent bias when computing the inverse covariance using the maximum likelihood estimator, and so we correct for this using the Hartlap correction $H$ \cite{Hartlap07}:
\begin{eqnarray}
    \boldsymbol{\Sigma}^{-1} \rightarrow H \boldsymbol{\Sigma}^{-1} \\
    H = \frac{N_H-D-2}{N_H-1},
\end{eqnarray}
where $D=9$ is the dimensionality of our summary statistic $\boldsymbol{s}$ (see below). Although we use 100,000 groups of slices, we only have $N_H=333$ independent observations. Given these values, the Hartlap correction is modest, $H\sim 0.95$.

When choosing a range of scales over which to measure each statistic, it generally makes sense to choose a large enough maximum scale such that all statistics can converge to their large-scale values. At the number densities considered here, we find that a maximum scale of $\sim 20$ cMpc (0.09 deg) at $z=5.7$ and $\sim 100$ cMpc (0.66 deg) at $z=6.6$ are reasonable choices. For example, in Figure \ref{fig:stats_comparison} we show the summary statistics at $z=6.6$ computed in 16 linearly-spaced bins between 1.5 and 105.0 cMpc. However, in the largest bins, the Betti-0 curve and VPF become concentrated near 1 and 0, respectively. To avoid issues with non-Gaussianity affecting our Fisher information results (see Appendix \ref{a:gaussianity}), when computing the Fisher information we discard the 5 largest bins from each statistic, and instead consider 10 linearly-spaced bins between 1.5 and 18.6 cMpc (0.41 and 5.13 arcmin; 1 and 12.4 voxels) at $z=5.7$ and 1.5 and 63.6 cMpc (0.60 and 25.23 arcmin; 1 and 42.4 voxels) at $z=6.6$. However, we find that whether or not we include these bins has a minimal 
effect on our results. Importantly, these scales are also well below the size of our simulation slices, which should minimize the impact of a limited simulation size on our results. Similarly, we find that the bin spacing we have chosen here ensures the convergence of our Fisher information calculation; doubling the number of bins over the same range, for example, has a minimal impact on our results.

Figure \ref{fig:fisher_standard} summarizes the main results of this work. In the top panels, we show the minimum uncertainties $\sigma_{\rm min} = 1/\sqrt{F}$ achievable by each statistic at $z=5.7$ and $z=6.6$ across a range of neutral fractions $x_{\rm HI}$. Solid colored lines show results for each summary statistic. Broadly, all statistics perform better at higher values of $x_{\rm HI}$, reflecting the trends seen in Figure \ref{fig:stats_comparison}. At $z=5.7$, the Betti-0 and VPF perform similarly to one another and both provide better constraints than the ACF, with the Betti-0 curve showing marginal improvement over the VPF at $x_{\rm HI} \gtrsim 0.15$. At $z=6.6$ and $x_{\rm HI} \lesssim 0.15$, all statistics perform similarly, while above this neutral fraction the VPF and ACF perform similarly, while the Betti-0 curve provides better constraints. Much tighter constraints are possible at $z=5.7$ likely owing to the higher density of sources. For further clarity, the middle panels show the fractional improvement of each statistic over the ACF, i.e., $(\sigma_{\rm min}^{\rm ACF}-\sigma_{\rm min})/\sigma_{\rm min}^{\rm ACF}$. At $z=5.7$, the Betti-0 curve and VPF provide 20-40\% smaller constraints than the ACF, with the Betti-0 outperforming the VPF marginally at $x_{\rm HI} > 0.15$. At $z=6.6$, the VFP performs similarly to the ACF, while the Betti-0 curve provides up to 20\% tighter constraints than the ACF at $x_{\rm HI} > 0.15$. As a visual aid, bottom panels show the inferred $x_{\rm HI}$ values with $1\sigma$ uncertainties given by $\sigma_{\rm min}$, assuming that the mean value is recovered exactly.

Our results comparing the VPF and ACF agree broadly with those of \cite{Gangolli21}, who find that the VPF outperforms the ACF in distinguishing a fully ionized Universe from one with $x_{\rm HI} = 0.3$ at $z=5.7$ in a SILVERRUSH.-like survey configuration but find that both perform similarly in distinguishing $x_{\rm HI}=0.0$ from $x_{\rm HI}=0.5$ at $z=6.6$. We note that we predict smaller uncertainties for the ACF than were computed in \citetalias{Umeda25}, which is likely for two reasons. First, we consider an average over multiple slices of \texttt{21cmFAST} to account for the reduced variance from a larger survey area, while \citetalias{Umeda25} considered the variance from individual slices. In addition, the variance from the Cram\'er-Rao bound represents a lower bound on the uncertainty when inferring $x_{\rm HI}$.

In Figure \ref{fig:fisher_standard} we also show uncertainties that are possible if pairs of statistics are constrained simultaneously, as well as if all three statistics are constrained. We compute these by simply appending the appropriate data vectors together when computing the Fisher information, e.g., $[\boldsymbol{s}_\beta, \boldsymbol{s}_f, \boldsymbol{s}_\omega]$. Importantly, constraints involving all three statistics perform better than any individual \textit{or} pair of statistics, meaning that the Betti-0 curve is not completely degenerate with the VPF or ACF and captures unique information compared with the other two statistics. The middle panels show that constraining all three statistics simultaneously improves constraints by between 40-60\% compared with the ACF. Compared with the best performing individual statistic (the Betti-0 curve), a simultaneous constraint improves the uncertainty by $\sim 30\%$. In Section \ref{ss:nondegenerate} we discuss why this is the case. 

The results in this section show that the Betti-0 curve can either be used on its own to provide similar or better constraints than the VPF or ACF, or it can be used in combination with the other statistics in order to improve constraints by a factor of 2 compared with the ACF alone. 

\subsection{Interlopers}\label{ss:interlopers}

\begin{figure*}
    \centering
    \includegraphics[width=1.\linewidth]{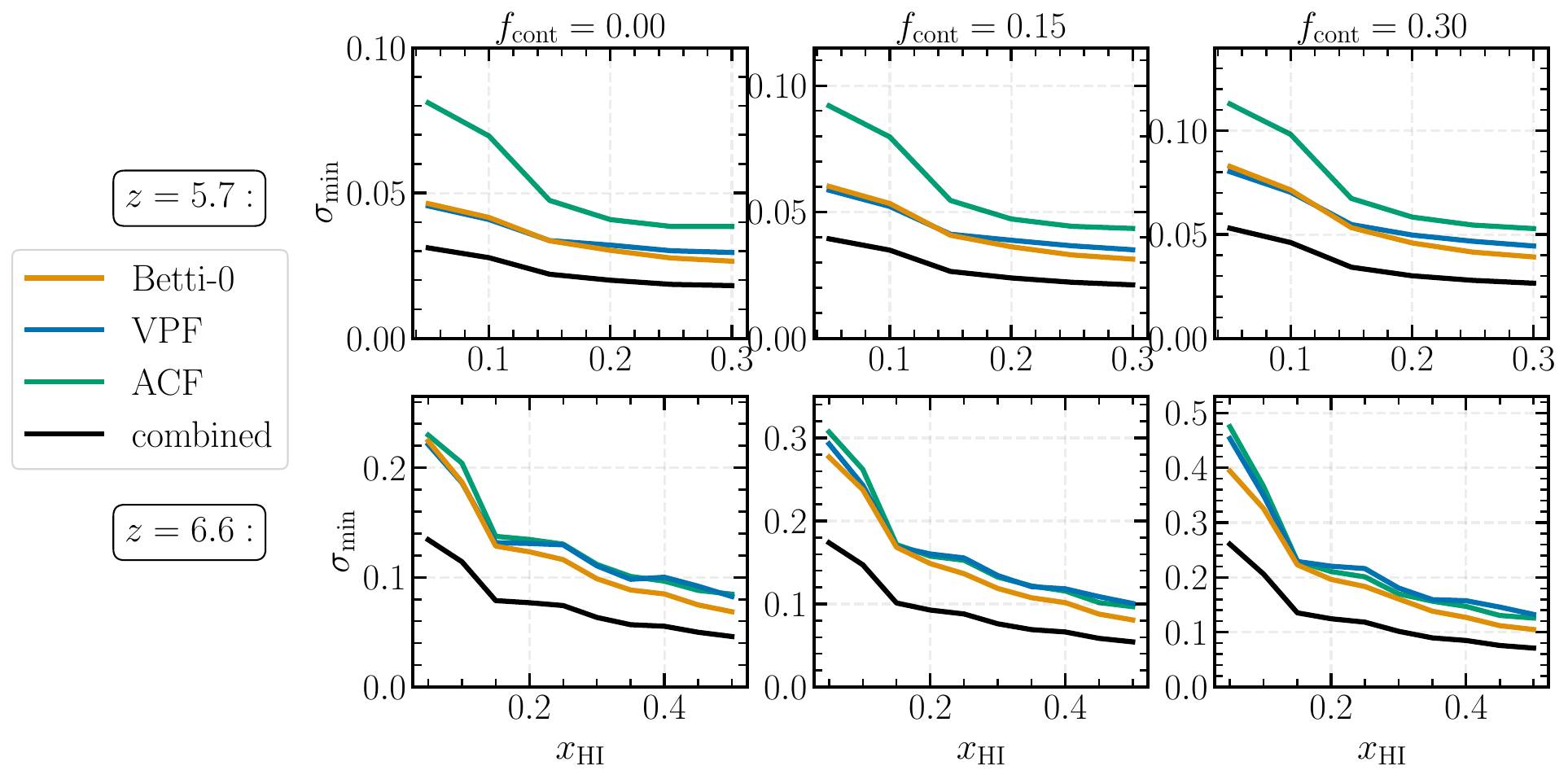}
    \caption{\textbf{The Betti-0 curve remains a promising summary statistic even at interloper fractions up to 30\%.} From left to right, panels show $\sigma_{\rm min}$ for all three summary statistics individually and combined at interloper fractions of $f_{\rm cont} = 0.00,$ 0.15, and 0.30 at $z=5.7$ (top row) and $z=6.6$ (bottom row). Note that the ordinate scale changes as the interloper fraction is increased --- while all statistics perform worse at higher interloper fractions, the relative performance of each remains similar. The left column shows the same results as Figure \ref{fig:fisher_standard}.} 
    \label{fig:interlopers}
\end{figure*}

Narrowband surveys are often susceptible to interlopers contaminating the LAE sample. As their correlations with the true LAE sample will not be modulated by the ionization morphology, contaminants will decrease the sensitivity of a summary statistic to $x_{\rm HI}$. It is therefore useful to understand how significant this effect will be. In this section we compare the effect of interlopers on each of the considered summary statistics.

As a spectroscopic follow-up on the SILVERRUSH. LAEs has not yet been completed, the interloper fraction has yet to be confirmed. \cite{Konno18} estimated the interloper fraction of an earlier SILVERRUSH. catalog with nearly identical selection criteria and spectroscopic follow up presented in \cite{Shibuya18, Shibuya18_ii} to be 14\% and 8\% at $z=5.7$ and 6.6 for all candidate LAEs, and \cite{Shibuya18_ii} estimated contamination fractions of $\sim33\%$ and $\sim 17\%$ for bright candidates with narrowband magnitudes $< 24$ (similar to our uniform selection; see Section \ref{ss:sim_obs}). However, \cite{Kikuta23} found that all bright contaminants reported in \cite{Shibuya18_ii} common to the SILVERRUSH. XIII. catalog (that considered in this work) were successfully omitted during the blind selection process. We nonetheless explore the effect of interlopers on the constraining power of the statistics; for a given interloper fraction, we remove the appropriate number of LAEs after selection and sampling and replace the same number with Poisson-distributed points. 

In Figure \ref{fig:interlopers}, we show results for all statistics with contamination fractions of $f_{\rm cont} = 0.00$, 0.15, and 0.30. Unsurprisingly, $\sigma_{\rm min}$ increases for all statistics with an increasing interloper fraction. However, all statistics respond similarly to interlopers, and maintain a similar relative performance (including the simultaneous fit of all three statistics), indicating that the Betti-0 curve remains a promising summary statistic compared with existing methods even the case of interlopers.

\section{Discussion}\label{s:discussion}
\subsection{Why are the statistics nondegenerate?}\label{ss:nondegenerate}
Although all statistics in this work are --- by design --- quite similar in that they somehow quantify the clustering of points, they are not degenerate with one another and each capture unique information, as we have shown in Section \ref{ss:fisher}. Since LAE clustering is highly non-Gaussian, and the ACF is inherently a Gaussian statistic, it is reassuring that the VPF and Betti-0 curve capture more information from the non-Gaussianity of the LAE distributions, but it is not obvious \textit{a priori} that the information captured by the ACF would not be a subset of this. In this section, we quantify the amount of information that each scale provides to the statistics. We find that each statistic is most sensitive to a different range of physical scales, and conclude that this is at least a partial explanation as to why the statistics are not degenerate.

\begin{figure}
    \centering
    \includegraphics[width=0.55\linewidth]{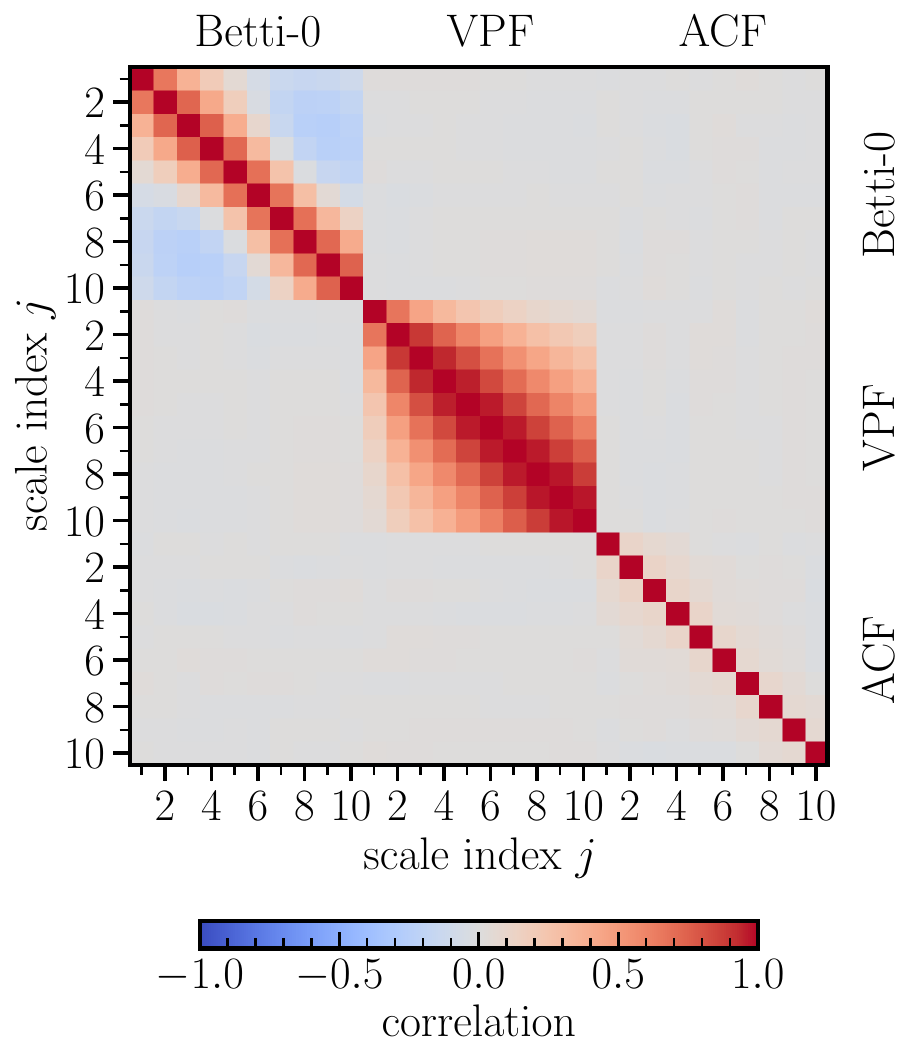}
    \caption{\textbf{Measurements of the ACF tend to be uncorrelated across scale, while those of the Betti-0 curve and VPF are highly correlated.} Colors show the values of the correlation matrix for the vector formed by appending the data vectors of all three statistics, $[\boldsymbol{s}_\beta, \boldsymbol{s}_f, \boldsymbol{s}_\omega]$. Diagonal blocks represent the correlations for each statistic with itself, and off-diagonal blocks show correlations between statistics. Off-diagonal blocks are shown, but all values are near zero.}
    \label{fig:corr_matrix}
\end{figure}

First, in Figure \ref{fig:corr_matrix}, we plot the correlation matrix of the vector formed by appending the data vectors of all three statistics, i.e., $[\boldsymbol{s}_\beta, \boldsymbol{s}_f, \boldsymbol{s}_\omega]$. The correlation matrix is a sort of normalized covariance matrix, whose components are
\begin{equation}
    C_{ij} = \Sigma_{ij} / \sigma_i \sigma_j,
\end{equation}
where $\Sigma_{ij}$ are the components of the covariance matrix and $\sigma_i = \sqrt{\Sigma_{ii}}$ are the standard deviations of the measurement in the $i$-th scale bin, i.e., the square root of the diagonal entries of the covariance matrix. Note that the correlation matrix is normalized to have values $[-1,1]$, and that all diagonal elements are equal to 1; we plot the correlation matrix rather than the covariance for exactly this reason, as otherwise the different ranges of values for each statistic would render the color bar unreadable. The diagonal blocks in Figure \ref{fig:corr_matrix} show the correlations of each statistic individually, and off-diagonal blocks show correlations of bins from different statistics.

Several things are apparent from this figure. First, the ACF tends to be uncorrelated across scales, as all off-diagonal elements within the ACF diagonal block are close to 0. On the other hand, the Betti-0 curve and VPF have strong correlations across scale, and subsequently non-zero off-diagonal elements. In particular, the Betti-0 curve exhibits positive correlations for nearby bins, while bins spaced further apart are anti-correlated, reflecting the trends discussed in figure \ref{fig:stats_example}. There are also no strong correlations between bins from different statistics, as all values in all off-diagonal blocks are near zero. We discuss potential non-Gaussianity arising from these correlations in Appendix \ref{a:gaussianity}.

Before continuing, we express the Fisher information as a sum of contributions of each of the eigenmodes $i$ of the covariance matrix, in terms of the eigenvalues $\lambda_i$ and eigenvectors $\boldsymbol{v_i}$ (see Appendix \ref{a:fisher}):
\begin{equation}\label{eq:fisher_decomposed}
    F = \sum_i \frac{(\boldsymbol{v}_i^T \boldsymbol{\mu}')^2}{\lambda_i} \equiv \sum_i \tilde{F}^{\rm mode}_i,
\end{equation}
where $\boldsymbol{\mu}' = d\boldsymbol{\mu}/d x_{\rm HI}$. Each term in this sum is the square of the projection of the derivative of the mean data vector onto the $i$-th eigenvector of the covariance, divided by the associated eigenvalue. These eigenvectors are the principal components of the covariance matrix; the eigenvectors of the covariance matrix form an orthonormal basis, where the largest eigenvalue characterizes the direction of largest variance, and each subsequent eigenvector points in the direction of largest variance in the orthogonal subspace which remains. In this way, the Fisher information quantifies changes in linear combinations of the data vector with $x_{\rm HI}$ normalized by the variance of that linear combination, so that larger changes with respect to the uncertainty contribute more information.

We denote by $(v_i)_j$ the component of the $i$-th eigenvector in the direction of the physical scale bin $j$. We quantify the amount of information that a scale $j$ contributes to each term in the sum \ref{eq:fisher_decomposed} by weighting each term by the square of this component: $\tilde{F}^{\rm scale}_{i,j} = \tilde{F}^{\rm mode}_i (v_i)_j^2$. Then, we define the information that a scale $j$ contributes to the total information as the sum across all eigenmodes $i$:
\begin{equation}\label{eq:fisher-scale}
    \tilde{F}^{\rm scale}_j = \sum_i \frac{(\boldsymbol{v}_i^T \boldsymbol{\mu}')^2}{\lambda_i} (v_i)_j^2.
\end{equation}
We note that the Fisher information inherently relies on the cross-terms of different scales, and so a true isolation of a given scale's contribution is not possible, nor is it necessarily desirable. All of the statistics encode multi-scale information, and Equation \ref{eq:fisher-scale} allows us to quantify how each statistic weights the different scales. Our chosen decomposition still relies on scales other than $j$ in each term through the dot product $(\boldsymbol{v}_i^T \boldsymbol{\mu}')$. Equation \ref{eq:fisher-scale} also has the desirable property of summing to the total Fisher information, since the eigenvectors are orthonormal, so that each term represents a fraction of the total information.

\begin{figure}
    \centering
    \includegraphics[width=0.575\linewidth]{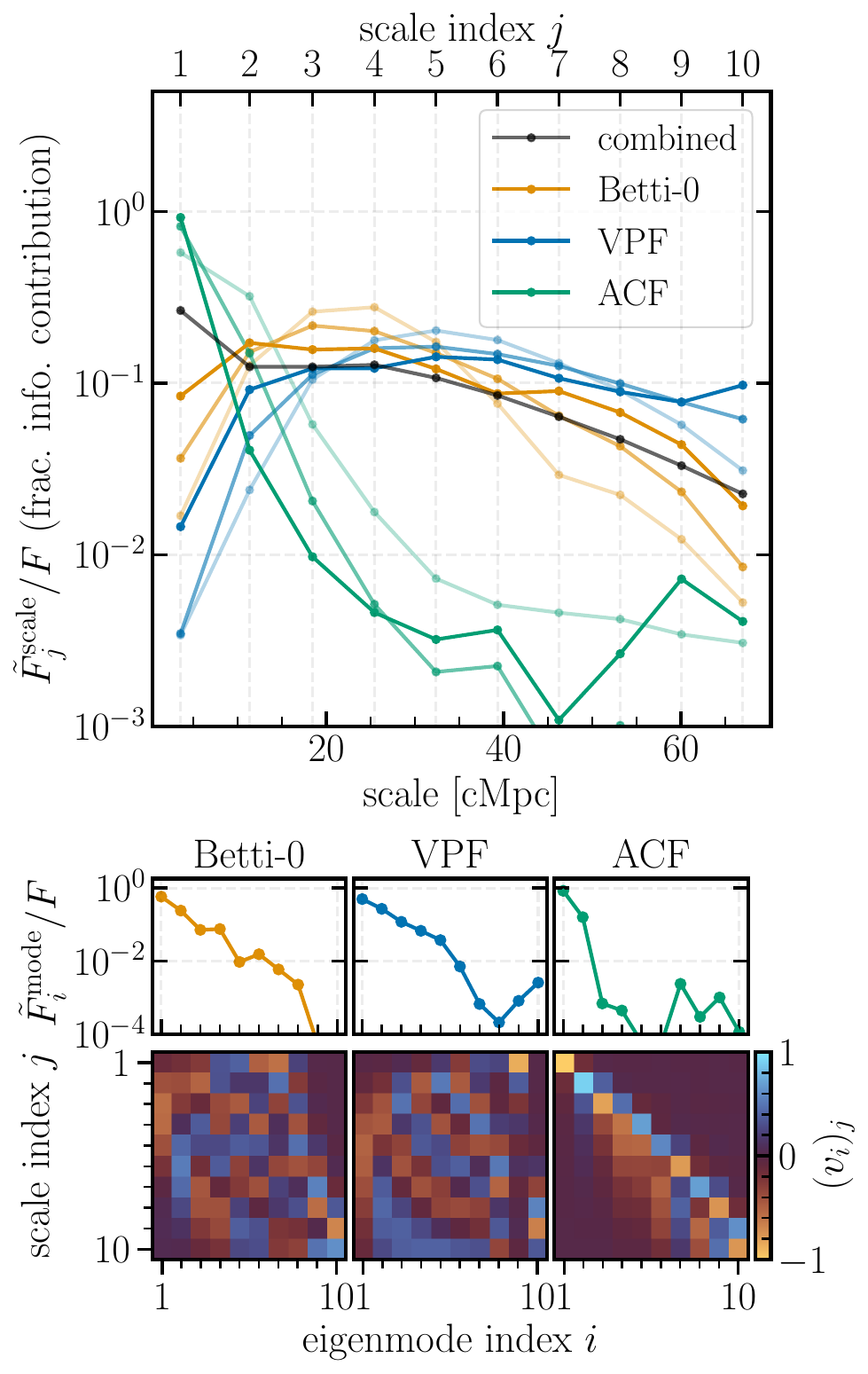}
    \caption{\textbf{The Betti-0 curve, the VPF, and the ACF are most sensitive to different physical scales, contributing to the fact that they capture complementary information.} The top panel shows the fractional contribution of each scale bin $j$ (labeled on the top) to the total information for the Betti-0, VPF, and ACF individually at $z=6.6$ and $x_{\rm HI} = [0.5,0.25,0.45]$ (shades, light to dark), and when fitting all three simultaneously (black) at $x_{\rm HI} = 0.25$ only. Middle panels show the fractional contribution of each eigenmode to the total information for each statistic. Bottom panels show the value of the component of the $i$-th eigenvector (columns) in the direction of the physical scale bin $j$ (rows), $(v_i)_j$. The middle and bottom panels show results for $x_{\rm HI} = 0.25$ only.}
    \label{fig:scale_contribution}
\end{figure}

In the top panel of Figure \ref{fig:scale_contribution}, we plot the fractional contribution of each scale $j$ to the total Fisher information, $\tilde{F}^{\rm scale}_j/F$ for each of the three statistics individually, as well as combined, at $z=6.6$ (with no contaminants) at various values of $x_{\rm HI}$. The scales which provide the most information to each statistic are clearly separated, with the ACF being most sensitive at the smallest scales, followed by the Betti-0 curve and finally the VPF. There is a mild evolution with $x_{\rm HI}$, but the same trends remain. This separation contributes to each statistic capturing different information from the LAE distributions and therefore at least partially explains why simultaneous constraints of all three statistics perform better than each individual or pair of statistics (Figure \ref{fig:fisher_standard}). 
For each individual statistic in Figure \ref{fig:scale_contribution}, we show the fractional contribution to its \textit{individual} total information. For the combined statistic, we consider the combined information. Because of this, we emphasize that Figure \ref{fig:scale_contribution} shows only the fractional contribution of scales to the information provided by each statistic (or the combination), and does not in any way provide a comparison of the total information of each, which is shown in Figure \ref{fig:fisher_standard}.

The bottom row of panels show, for $x_{\rm HI} = 0.25$, the eigenvectors of each mode $i$ in columns, and the component of each eigenvector in the direction of scale $j$ in rows, so that the color indicates the value of the component $(v_i)_j$. The middle panels show the fractional contribution of each eigenmode $i$ to the total information $\tilde{F}_i/F$. In all three statistics, the first few eigenmodes dominate the information. Looking more closely at the ACF, we can see that these first few eigenvectors are closely aligned with the original basis of scales $j$, since the largest of their components lie along the diagonal. This is because the ACF tends to be uncorrelated across scales (see Figure \ref{fig:corr_matrix}). As a specific example, the first eigenvector~$i=1$ (the left hand column) shows only a significant component in the first scale bin $j=1$, indicating that this scale provides most of the information in this eigenmode. Looking at the fractional contribution of this eigenmode $\tilde{F}^{\rm mode}_{i=1}/F$, we can see that it provides a large fraction of the total information. Together, this implies that the first scale bin provides a significant fraction of the total information for the ACF, which is reflected in the top panel.

The Betti-0 and VPF eigenvectors are more complex, and the dominant modes tend to be combinations of multiple scales, reflecting the fact that measurements of these statistics across scales are highly correlated (see Figure \ref{fig:corr_matrix}). The Betti-0 curve and ACF contain little information at the smallest scales due to their construction, as both begin at the same value at a scale of 0 regardless of the clustering of points. This forces the smallest bins to be more similar across $x_{\rm HI}$, reducing their information content.

Looking at the fractional information contribution across scales when simultaneously fitting all three statistics (top panel, black curve), it is clear that combining the statistics distributes the information across a wider range of scales compared with any statistic individually, which subsequently increases the overall information content (as shown in Figure \ref{fig:fisher_standard}).

The trends discussed here also explain why comparisons like Figure \ref{fig:stats_comparison} can be misleading in terms of the performance of statistics for parameter inference. Because eigenvectors of the ACF tend to be aligned with the original basis (the physical scales over which it is measured), this means that the components of the projection of the data onto this basis (Equation \ref{eq:fisher_decomposed}) are approximately the original data components themselves, i.e., $\boldsymbol{v}_i^T \boldsymbol{\mu}' \approx \mu'_i$, and the eigenvalues are approximately the uncertainties of each individual data point, i.e., $\lambda_i \approx \sigma_i$. In such a case, the separations reported in Figure \ref{fig:stats_comparison} properly represent the sensitivity of the statistic to the parameter in question. However, because the Betti-0 curve and ACF are highly correlated across scales, their eigenvectors are linear combinations of multiple scales, and differences as shown in Figure \ref{fig:stats_comparison} do not accurately portray their sensitivities. If instead one treated their components as independent, like the ACF, one would overestimate their constraining power.

\subsection{Sensitivity to astrophysics}\label{ss:sensitivity_astro}

Statistics used to constrain $x_{\rm HI}$ should be more sensitive to the ionized bubble morphology than to uncertainties in the underlying astrophysics, as discussed in \cite{Mesinger08_early}. Throughout this work, we have used only a single set of astrophysical parameters and therefore the sensitivities reported are subject only to cosmic variance uncertainties. Ideally, one would marginalize over a variety of astrophysical parameter choices when fitting for $x_{\rm HI}$, but this is beyond the scope of the current work. Instead, we choose a simple approximation to demonstrate the robustness of the Betti-0 curve to astrophysical uncertainties, at least compared with the VPF and ACF.

To do this, we marginalize over a range of choices for $L_{\rm min}$ but keep the number of LAEs fixed by randomly sampling LAEs above the threshold luminosity. This implicitly changes the assumed mapping between halo mass and $M_{\rm UV}$, on which our $\lya$ luminosities are dependent, while keeping the ionization morphology unchanged. In detail, we consider $z=6.6$ and set $\log_{10} L_{\rm min}$ to 10 linearly-spaced values between 42.5 and 42.95 erg s$^{-1}$. The lower limit represents a factor of 2 change in the threshold luminosity from the fiducial value of 42.8, and beyond the upper limit we do not recover enough LAEs for our normalization. Our procedure is similar to before, except we only compute 20 random samples for each slice, and form 200,000 groups of 6 slices each, where for each group we also randomly choose a value for $L_{\rm min}$.

\begin{figure}
    \centering
    \includegraphics[width=0.5\linewidth]{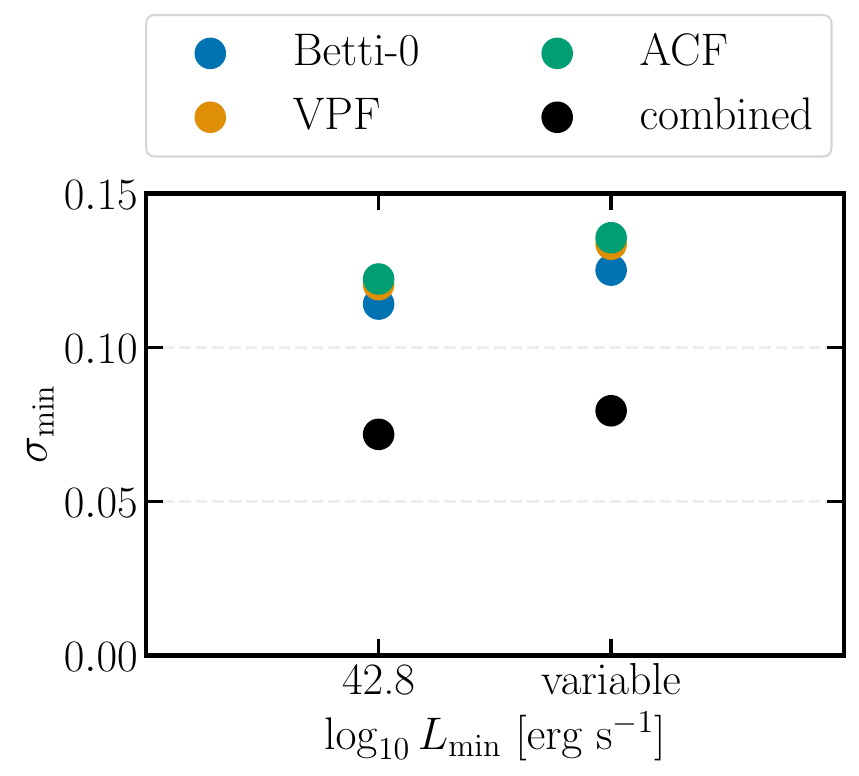}
    \caption{\textbf{The Betti-0 curve is similarly robust to uncertainties in the astrophysics as the VPF and ACF.} Points show $\sigma_{\rm min}$ for all statistics at $z=6.6$ and $x_{\rm HI} = 0.25$ for the fiducial threshold luminosity of $\log_{10} L_{\rm min} = 42.8$~erg~s$^{-1}$ and for a marginalization over the threshold luminosity between $42.5 \leq \log_{10} L_{\rm min}/\text{erg s}^{-1} \leq 42.95$ as an approximation for a marginalization over astrophysical parameters. The fractional change in the minimum uncertainty for the Betti-0 curve, VPF, ACF, and combination is $(\sigma_{\rm min}^{42.8} - \sigma_{\rm min}^{\rm var})/\sigma_{\rm min}^{42.8} = [-0.09,-0.08,-0.09,-0.08]$.}
    \label{fig:variable_Lmin}
\end{figure}

In Figure \ref{fig:variable_Lmin}, we show $\sigma_{\rm min}$ for $\log_{10} L_{\rm min} = 42.8$ erg s$^{-1}$ and for the marginalization across $L_{\rm min}$. We show results at $z=6.6$ and $x_{\rm HI} = 0.25$. As expected, when marginalizing over $L_{\rm min}$, all statistics become less constraining, though the change is modest and, importantly, similar for all statistics, including the Betti-0 curve. The fractional change in the minimum uncertainty for the Betti-0 curve, VPF, ACF, and combination is $(\sigma_{\rm min}^{42.8} - \sigma_{\rm min}^{\rm var})/\sigma_{\rm min}^{42.8} = [-0.09,-0.08,-0.09,-0.08]$. This indicates that the Betti-0 curve is as robust to changes in the underlying astrophysics as the VPF and ACF.

\section{Conclusion}\label{s:conclusion}

In this work we have introduced persistent homology, and in particular the Betti-0 curve, as a new method for characterizing the clustering of LAEs and for use as a summary statistic in parameter inference of reionization. Using a suite of simulations from \texttt{21cmFASTv4}, run using the same astrophysical parameters and different starting seeds, and an empirically-calibrated model for assigning $\lya$ luminosities to galaxies, we forward model LAE observations similar to those from the newest SILVERRUSH. catalogs. Using the Fisher information to compute the minimum uncertainties achievable on the global ionized fraction $x_{\rm HI}$, we show that, when applied to these observations, the Betti-0 curve shows promise as a summary statistic. In particular, we find:
\begin{enumerate}
    \item The Betti-0 curve provides constraints on $x_{\rm HI}$ which are as small as, or better than, either the VPF or ACF. Accounting for uncertainties only due to cosmic variance, we predict minimum uncertainties on constraints of $x_{\rm HI}$ using the Betti-0, VPF, and ACF of  $\sigma_{\rm min} \sim$ 0.05, 0.05, and 0.08, respectively at $x_{\rm HI} = 0.05$ and $z=5.7$ and of $\sigma_{\rm min} \sim$ 0.12, 0.13, and 0.13 at $x_{\rm HI} = 0.20$ and $z=6.6$ (Section \ref{ss:fisher}).
    \item Simultaneously constraining all three statistics improves constraints over the ACF by a factor of $\sim 2$, with predicted minimum uncertainties of $\sigma_{\rm min} \sim$ 0.03 and 0.07 at $x_{\rm HI} = 0.05$ and $z=5.7$ and $x_{\rm HI} = 0.20$ and $z=6.6$, respectively (Section \ref{ss:fisher}).
    \item All three statistics respond similarly to contaminants, and the Betti-0 curve maintains the same relative performance compared with the VPF and ACF in the case of contaminants (Section \ref{ss:interlopers}).
    \item Each statistic is most sensitive to a different range of physical scales, and therefore captures complementary information to the other two. This at least partially explains why fitting all three simultaneously improves constraints (Section \ref{ss:nondegenerate}).
    \item An approximate test indicates that the Betti-0 curve is as robust to uncertainties in the underlying astrophysics as the VPF and ACF (Section \ref{ss:sensitivity_astro}).
\end{enumerate}

For the sake of a practical comparison, we have focused our analysis in this work within the context of the SILVERRUSH. survey. However, we expect our results to remain broadly consistent for any similar wide-area LAE survey, and we therefore conclude that the Betti-0 curve should be considered a promising summary statistic when performing inferences of reionization using LAE observations in such surveys, particularly for existing wide-field surveys such as SILVERRUSH. and for potential future surveys with the Nancy Grace Roman Telescope. In surveys where spectroscopic redshifts are available (such as with Roman or the Subaru Prime Focus Spectrograph), the Betti-0 curve can also be applied in 3-dimensions. In addition, we reiterate that persistent homology is a rich formalism, and the analysis in this work considers only the simplest possible summary within this framework. Because of this, we also advocate for the continued study of potentially more complex summaries based on persistent homology for characterizing LAE clustering. 

\begin{acknowledgments}
M.M.W. thanks E.F. Bunn for helpful discussions regarding Fisher information and linear algebra. M.M.W. and S.R.F. were supported by NASA through award 80NSSC22K0818 and by the National Science Foundation through award AST-2510939. N.T. acknowledges the Italian Ministerial grant PNRR from National Centre for HPC, Big Data and Quantum Computing CUP E53C22000790001, Spoke 3. N.T. acknowledges CINECA award under the ISCRA initiative for providing us access to the LEONARDO supercomputer (IsB30\_IC-diff, project no. HP10B1D1F2, PI: Triantafyllou). N.T., S.G.-H., and A.M. gratefully acknowledge the computational resources of the HPC center at SNS. M.O. acknowledges the supports from the World Premier International Research Center Initiative (WPI Initiative), MEXT, Japan, the joint research program of the Institute for Cosmic Ray Research (ICRR), the University of Tokyo, and KAKENHI (21H04467, 25H00674) through JSPS. The authors acknowledge the use of ChatGPT to assist with language editing and code development and de-bugging. All AI-assisted outputs were carefully reviewed and validated by the authors. The authors take full responsibility for all analyses, interpretations, and conclusions presented in this work. 

This work has made extensive use of the SAO/NASA Astrophysics Data System
(\url{http://ui.adsabs.harvard.edu}), the arXiv e-Print service (\url{http://arxiv.org}), as well as the following softwares: \texttt{matplotlib} \cite{Matplotlib}, \texttt{numpy} \cite{numpy}, \texttt{astropy} \cite{Astropy}, and \texttt{scipy} \cite{Scipy}, \texttt{Ripser} \cite{Ripser}, and \texttt{Ripser.py} \cite{ripser_py}, ChatGPT \cite{ChatGPT}.
\end{acknowledgments}

\appendix

\section{Gaussianity of the statistics}\label{a:gaussianity}

\begin{figure}
    \centering
    \includegraphics[width=1.\linewidth]{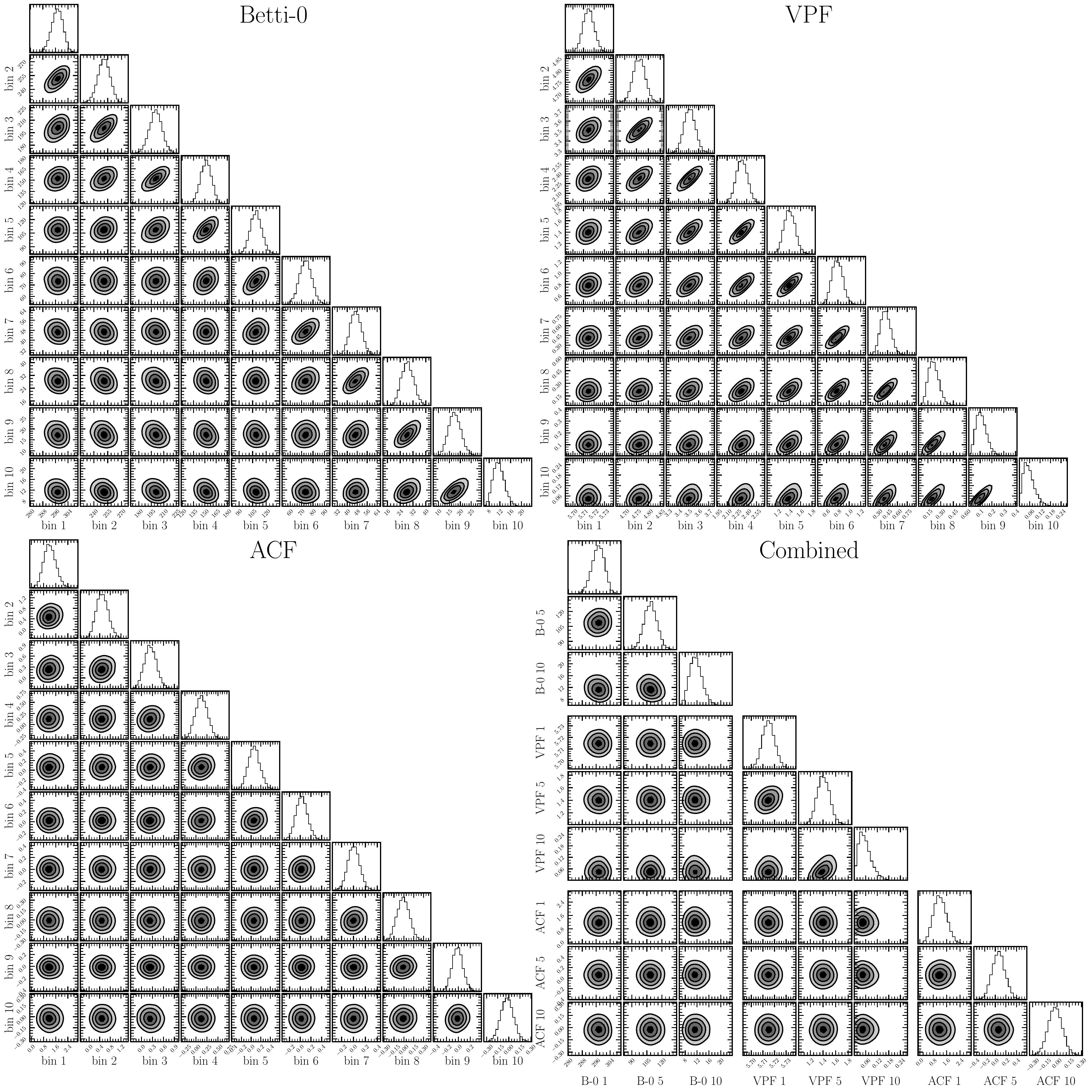}
    \caption{\textbf{All statistics do not exhibit obvious non-Gaussianities at the two-point level.} Diagonal panels show 1-dimensional histograms of the values for each bin of each statistic across a single set of 2,000 iterations from the simulations at $z=6.6$ and $x_{\rm HI} = 0.25$. Off-diagonal panels show the pairwise distributions of each bin. The bottom right corner plot shows the same for the combined data vector.}
    \label{fig:corner_plot}
\end{figure}

The Betti-0 curve and VPF are highly correlated across scales, leading to the potential for them to non-Gaussian. In figure \ref{fig:corner_plot}, we show a corner plot of the distributions of each individual bin for all three statistics, as well as their pairwise distributions at $z=6.6$ and $x_{\rm HI} = 0.25$. In addition, we show the same distributions for a selection of bins from the combined statistic. Unsurprisingly, nearby bins tend to be positively correlated for both the Betti-0 curve and ACF. For the Betti-0 curve, bins spaced further apart tend to be slightly anti-correlated, reflecting the turnover discussed in figure \ref{fig:stats_example}. Bins in the ACF do not exhibit strong correlations. The corner plot for the combined data vector shows the same trends within each individual statistic, and shows no strong correlations in bins across different statistics. The trends mentioned here follow those discussed in Section \ref{s:discussion}. In all cases, the distributions do not visually exhibit any obvious non-Gaussianity, and so we do not expect the true information to deviate significantly from that measured using Equation \ref{eq:fisher}. However, potential non-Gaussianity does motivate the use of more sophisticated methods for inferences using these statistics, such as simulation-based inference, which requires no assumption of Gaussianity.

\section{Fisher information per eigenmode of the covariance matrix}\label{a:fisher}

We can diagonalize the covariance matrix as $\boldsymbol{\Sigma} = \boldsymbol{V \Lambda V}^T$ and thus we have $\boldsymbol{\Sigma}^{-1} = \boldsymbol{V \Lambda}^{-1} \boldsymbol{V}^T$, where the columns of $\boldsymbol{V}$ are comprised of the orthonormal eigenvectors of $\boldsymbol{\Sigma}$, $\boldsymbol{v}_i$, and $\boldsymbol{\Lambda}$ is a diagonal matrix whose entries are the associated eigenvalues (i.e., $\lambda_{ij} = 0 $ unless $i=j$ in which case $\lambda_{ii} \equiv \lambda_i $). Denoting the derivatives with respect to $x_{\rm HI}$ using prime notation, we can rewrite the Fisher information as

\begin{equation}
    F = (\boldsymbol{\mu}'^T \boldsymbol{V}) \boldsymbol{\Lambda}^{-1} (\boldsymbol{V}^T \boldsymbol{\mu}').
\end{equation}
For clarity, we have included suggestive brackets; $\boldsymbol{V}^T \boldsymbol{\mu}'$ is the projection of the derivative of the mean of our data vector $\boldsymbol{\mu}'$ onto the basis formed by the eigenvectors.
Denoting the components of $\boldsymbol{\mu}'$ in the eigenbasis
by $(\boldsymbol{V}^T \boldsymbol{\mu}')_i = \boldsymbol{v}_i^T \boldsymbol{\mu}'$,
and using the fact that $\boldsymbol{\Lambda}$ is diagonal, we have

\begin{align}
\begin{split}
    F = ( \boldsymbol{\mu}'^T \boldsymbol{V}) \boldsymbol{\Lambda}^{-1} (\boldsymbol{V}^T \boldsymbol{\mu}') &= (\boldsymbol{V}^T \boldsymbol{\mu}' )^T \left[ \boldsymbol{\Lambda}^{-1} (\boldsymbol{V}^T \boldsymbol{\mu}') \right] \\
    &= \sum_i (\boldsymbol{V}^T \boldsymbol{\mu}' )_i \left[ \boldsymbol{\Lambda}^{-1} (\boldsymbol{V}^T \boldsymbol{\mu}') \right]_i \\
    &= \sum_i (\boldsymbol{V}^T \boldsymbol{\mu}' )_i \left[ \sum_j \lambda^{-1}_{ij} (\boldsymbol{V}^T \boldsymbol{\mu}')_j \right] \\
    &= \sum_i (\boldsymbol{V}^T \boldsymbol{\mu}' )_i \left[ \lambda^{-1}_{i} (\boldsymbol{V}^T \boldsymbol{\mu}')_i \right] \\
    &= \sum_i  \frac{ (\boldsymbol{v}^T_i \boldsymbol{\mu}')^2}{\lambda_{i}}.
\end{split}
\end{align}

\bibliography{bib}{}
\bibliographystyle{JHEP.bst}

\end{document}